\documentclass[superscriptaddress,amsmath,amssymb,aps,reprint]{revtex4-2}

\usepackage{graphicx}
\usepackage{dcolumn}
\usepackage{bm}
\usepackage{float}
\usepackage{subfigure}
\usepackage{hyperref}
\usepackage{upgreek}
\usepackage{siunitx}
\usepackage{soul,color,xcolor}
\soulregister{\cite}7
\soulregister{\citep}7
\soulregister{\citet}7
\soulregister{\ref}7
\soulregister{\pageref}7
\allowdisplaybreaks[2]
\hypersetup{colorlinks=true,linkcolor=black,filecolor=black,urlcolor=blue,citecolor=blue}

\begin{document}

\preprint{APS/123-QED}

\title{Thermal conduction force under standing and quasi-standing temperature field}

\author{Haohan Tan}
\affiliation{Department of Physics, State Key Laboratory of Surface Physics, and Key Laboratory of Micro and Nano Photonic Structures (MOE), Fudan University, Shanghai 200438, China}

\author{Yuqian Zhao}
\affiliation{Department of Physics, State Key Laboratory of Surface Physics, and Key Laboratory of Micro and Nano Photonic Structures (MOE), Fudan University, Shanghai 200438, China}

\author{Jiping Huang}
\thanks{jphuang@fudan.edu.cn}
\affiliation{Department of Physics, State Key Laboratory of Surface Physics, and Key Laboratory of Micro and Nano Photonic Structures (MOE), Fudan University, Shanghai 200438, China}

\date{\today}

\begin{abstract}
  Thermal conduction force plays a crucial role in manipulating the local thermal conductivity of crystals. However, due to the diffusive nature of thermal conduction, investigating the force effect is challenging. Recently, researchers have explored the force effect based on the wave-like behavior of thermal conduction, specifically second sound. However, their focus has been primarily on the progressive case, neglecting the more complex standing temperature field case. Additionally, establishing a connection between the results obtained from the progressive case and the standing case poses a challenging problem. In this study, we investigate the force effect of standing and quasi-standing temperature fields, revealing distinct characteristics of thermal conduction force. Moreover, we establish a link between the progressive and standing cases through the quasi-standing case. Our findings pave the way for research in more intricate scenarios and provide an additional degree of freedom for manipulating the local thermal conductivity of dielectric crystals.
\end{abstract}

\maketitle

\section{Introduction}

Thermal conduction is a research field that has garnered significant attention, both from a macroscopic and microscopic perspective~\cite{zhou2023adaptive,xu2023giant,yang2023controlling,xu2022diffusive,xu2022thermal,jin2023tunable,zhang2023diffusion,li2021transforming,palla2020stochastic,lepri2003thermal,dhar2008heat,yang2017full}. In the thermal metamaterial field, researchers have introduced an array of devices with novel functionalities, encompassing thermal cloaks~\cite{xu2020active,xu2021geometric,xu2020transformation,xu2020transformation1}, concentrators~\cite{dai2018transient}, illusion~\cite{zhu2015converting,yang2019thermal}, transparency~\cite{xu2019thermal}, and sensors~\cite{jin2020making}. However, the force effect associated with this heat transport mechanism has received limited attention. In this regard, the thermal conduction force between the liquid-liquid and liquid-solid phases has been theoretically explained and experimentally verified by considering the coupling of momentum and flux~\cite{gaeta1969radiation,gaeta1991radiation,albanese1997experimental}. For the solid-solid case, Tan et al. proposed the theory of second sound radiation force (SSRF) based on the wave-like nature of thermal conduction in dielectric crystals~\cite{tan2023tunable,huberman2019observation,ding2022observation,beardo2021observation,jackson1970second,narayanamurti1972observation,osborne1951second}. Their work not only confirmed the existence of SSRF but also demonstrated its tunability by manipulating the incident wave. This research challenges the conventional notion that a constant temperature gradient is necessary to induce thermal conduction force. Moreover, SSRF provides a mechanism for manipulating local thermal conductivity~\cite{callaway1959model,guyer1966thermal}.

However, Tan et al. only investigated the case of a single incident second sound~\cite{tan2023tunable}. To further advance the research, it is necessary to explore the scenario involving multiple incident second sounds. When two second sounds with equal amplitudes counter-propagate, they create a standing temperature field, while different amplitudes result in a quasi-standing temperature field. Consequently, the thermal conduction force becomes more complex. Firstly, it is unknown whether the force exhibits distinct behavior compared to the progressive case under these two scenarios. Secondly, it remains uncertain whether changing the force direction, as previously observed, holds the same significance. Furthermore, we are particularly interested in determining if there is a relationship between the results obtained from these different cases.
\begin{figure}
  \includegraphics[width=1.0\linewidth]{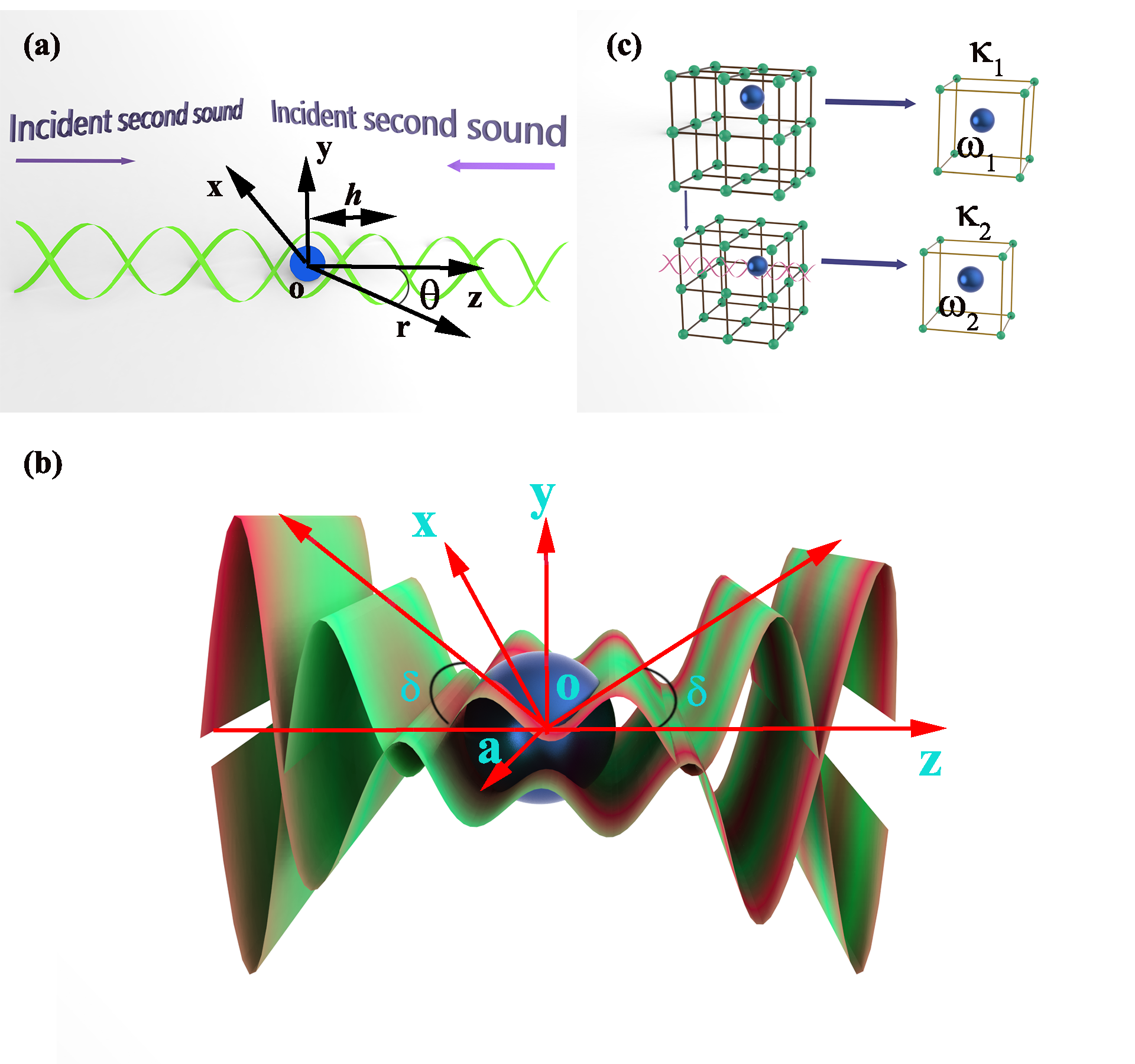}
	\caption{Schematic of thermal conduction force under standing and quasi-standing temperature field. (a) An adiabatic spherical particle in the presence of  plane standing temperature field is depicted. The particle center is the origin of the Cartesian coordinate system $(x,y,z)$, and $\theta$ denotes the polar angle. (b) A particle with radius $a$ in the presence of Bessel standing second sound is shown, with $\delta$ representing the half cone angle. (c) The local thermal conductivity of dielectric crystals can be tuned by manipulating the interaction between the impurity particle (the blue sphere) and the second sound (the pink wave). $\omega_{\rm 1}$ and $\omega_{\rm 2}$ are the vibrating angular frequencies of the impurity particle before and after interacting with the second sound, and $\kappa_{\rm 1}$ and $\kappa_{\rm 2}$ are the corresponding local thermal conductivities of the crystal. The region containing the impurity particle is zoomed in for clarity.}
	\label{f1}
\end{figure}
In this study, we theoretically investigate the thermal conduction force exerted on stationary and adiabatic impurity particles in dielectric crystals when two counter-propagating second sounds generate standing and quasi-standing temperature fields. We consider various cases, including the plane, zeroth-order Bessel, and high-order Bessel standing cases. Additionally, we examine the corresponding quasi-standing cases. Interestingly, we discover that the force direction can be reversed, similar to the progressive case. However, due to the absence of a specific propagating direction for the standing and quasi-standing temperature fields, their physical implications differ. The most significant finding is that the results obtained from the quasi-standing case establish a connection between the results of the standing and progressive cases. These findings not only enhance the theoretical understanding of thermal conduction force but also provide an additional degree of freedom for manipulating the local thermal conductivity of dielectric crystals.

\section{General theory for calculating thermal conduction force}
To investigate the standing and quasi-standing temperature field's impact on the SSRF, we utilize the non-attention model as progressive case. Initially, we assume that the dielectric crystal's temperature is approximately $10$~K, ensuring that phonon transport operates within the hydrodynamic regime. Subsequently, we assume that the resistive process can be neglected, implying an infinite resistive relaxation time, $\tau_{\rm R}\to \infty$. Additionally, we consider the normal relaxation time to be sufficiently short. Consequently, we can assume that the phonon system can be described by the local equilibrium function
\begin{equation}
     f(\boldsymbol{k},\boldsymbol{r},t)=\frac{1}{e^{\beta(\boldsymbol{r},t)(\hbar\omega_{k}-\hbar\boldsymbol{k}\cdot\boldsymbol{u}(\boldsymbol{r},t))}-1},
  \end{equation}
  where $\hbar$, $\omega_{k}$, $\bm{k}$, and $\bm{u}(\bm{r},t)$ are the reduced Planck constant, phonon angular frequency, phonon wave vector, and phonon drifting velocity, respectively. $\beta(\bm{r},t)=1/(k_{\rm B}T(\bm{r},t))$, where $k_{\rm B}$ is the Boltzmann constant, and $T(\bm{r},t)$ is the local temperature of the crystal. $\bm{r}$ is the position vector, and $t$ represents time.
  In our investigation of the force effect of second sound, the size of the impurity particle has an impact on the resistive relaxation time. However, our focus lies within the regime where $\tau_{\rm N} << \tau_{\rm R}$. Here, $\tau_{\rm N}$ represents the normal relaxation time. Under this condition, the local equilibrium assumption remains valid, even as the radius of the impurity particle, denoted as $a$, varies. Through the application of momentum and energy conservation principles, as well as the utilization of the Debye model to characterize the phonon spectrum, we are able to provide evidence supporting the propagation of the temperature field as a wave phenomenon known as second sound~\cite{ward1951velocity,Dingle1952Derivation,kwok1967dispersion,sussmann1963thermal,beck1974phonon,chester1963second,prohofsky1964second,li2022second}.

According to the momentum conservation law, we can define the SSRF as the progressive case.
\begin{equation}
    F=-\iint \left \langle {\it\Pi^{\rm '}}\right \rangle \mathrm{d}S,
  \label{F1}
  \end{equation}
  where $\it\Pi_{\rm ij}^{\rm '}=\sum_{k}\hbar k_{\rm i} v_{k_{\rm j}}\left(f-f_{\rm 0}\right)$ is the momentum flux of the second sound, and $v_{k_{\rm j}}=\partial\omega_{k}/{\partial k_{\rm j}}$ is the phonon group velocity component. The integration is over the particle surface $S$.
  In this context, the term $f-f_{0}$ represents the deviation of the phonon system from its equilibrium state, which is characterized by $f_{0}$. The equilibrium state of the system is given by the equation:

  \begin{equation}
	  f_{\rm 0}(\boldsymbol{k})=\frac{1}{e^{\beta_{\rm 0}\hbar\omega_{k}}-1},
  \end{equation}
  where $\beta_0=1/k_BT_0$ and $T_0$ represents the background temperature. In the equilibrium state ($f=f_{0}$), there is no SSRF present.
  
  For the sake of simplicity, we assume that one incident second sound propagates along the $+z$ direction, while the other propagates along the $-z$ direction. Under this assumption, Equation (\ref{F1}) can be further simplified as follows:\begin{eqnarray}
    F=-\iint \sum_{k}\hbar k_{\rm z} v_{k_{\rm z}}\left(f-f_{\rm 0}\right)\mathrm{d}S.
    \label{eq12}
  \end{eqnarray}
 After further derivation, we can obtain the expression for the SSRF under standing and quasi-standing case.
 \begin{align}
	F&=-\frac{1}{2}\times\nonumber\\
	&\iint \left(\frac{2\pi^{\rm 2}}{9\beta_{\rm 0}^{\rm 4}(\hbar c)^{\rm 3}}\left\langle \left( \frac{T_{\rm 1}^{\rm '}}{T_{\rm 0}}\right)^{\rm 2}\right\rangle +\frac{\pi^{\rm 2}}{15\beta_{\rm 0}^{\rm 4}\hbar^{\rm 3} c^{\rm 5}} \left\langle \left|u\right|^{\rm 2} \right\rangle \right)\mathrm{d}S,
	\label{eq14}
\end{align}
Here, $c$ represents the modulus of the phonon group velocity. It is noteworthy that the expression derived above for the SSRF is identical to the one obtained in the progressive case. However, in this context, $T_{1}^{'}$ represents the sum of two incident second sounds and their corresponding scattering field. When calculating the acoustic radiation force in the acoustic field, two methods are commonly employed: the angular-spectrum method and the multipole-expansion based method. Extensive research has demonstrated the equivalence of these two approaches \cite{baudoin2020acoustic, sapozhnikov2013radiation, gong2020acoustic, silva2011expression, baresch2013three, silva2012radiation, gong2019reversals, gong2021equivalence}. In the subsequent analysis, we will utilize the multipole-expansion based method for direct calculations.

\begin{figure*}[!ht]
	\includegraphics[width=1.0\linewidth]{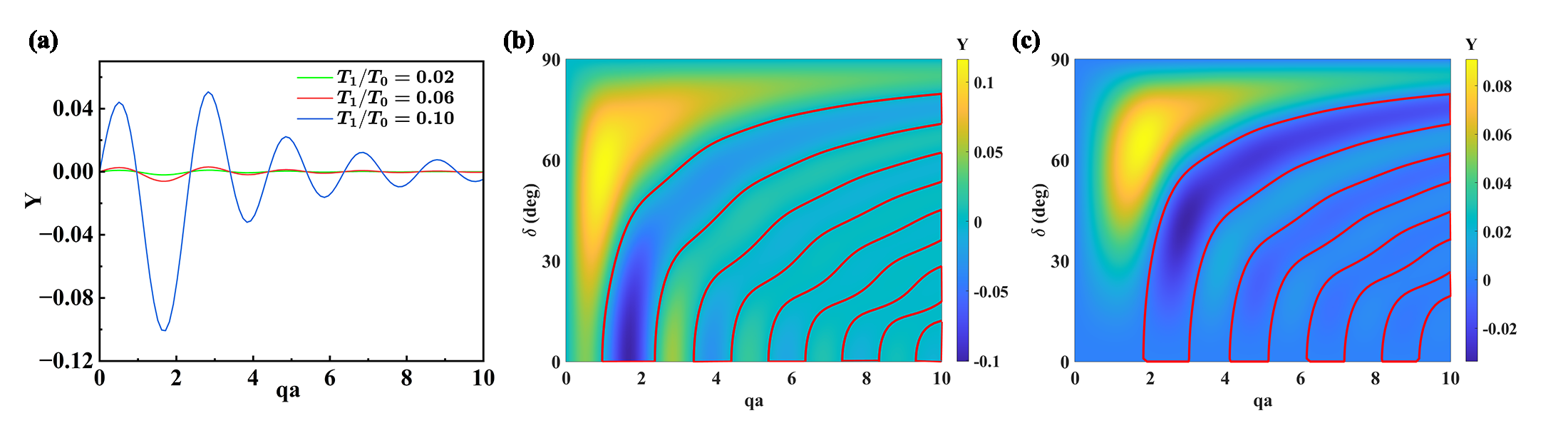}
	\caption{Numerical results of the reduced second sound radiation force $Y_{st}$ under the conditions of (a) plane standing second sound as a function of $qa$ (where $q$ and $a$ are the second sound wave number and the particle radius) with the temperature ratio $T_{\rm 1}/T_{\rm 0}=0.02, 0.06, 0.10$ (where $T_1$ is the second sound amplitude, and $T_0$ is the background temperature), (b) zeroth-order Bessel standing second sound as a function of $qa$ and $\delta$ with $T_{\rm 1}/T_{\rm 0}=0.10$, and (c) first-order Bessel standing second sound as a function of $qa$ and $\delta$ with $T_{\rm 1}/T_{\rm 0}=0.10$. The area enclosed by the red line is where $Y_{st}<0$.}
	\label{img11}
\end{figure*}
\section{force effects with different standing and quasi-standing temperature fields}
Initially, we examine the SSRF under a plane standing temperature field. In this scenario, the incident second sound can be mathematically represented as $T_{\rm inc}=T_1e^{-i\omega t}\{e^{iq(z+h)}+e^{-iq(z+h)}\}$, where $T_{1}$, $\omega$, and $q$ denote the amplitude, angular frequency, and wave number of the second sound, respectively. Here, $h$ represents the distance in the $z$-direction from the center of the impurity to the nearest velocity antinode, as is shown in Fig.~\ref{f1}(a). Additionally, $i=\sqrt{-1}$, and $t$ signifies time. By expressing the incident second sound in spherical coordinates, we can derive the following expression:
\begin{align}
  T_{\rm inc}=T_{\rm 1}\sum_{n=0}^{\rm \infty}(2n+1)\Lambda_{\rm n}i^{\rm n}j_{\rm n}(qr)P_{\rm n}(\cos\theta)e^{-i\omega t},
\end{align}
 where $\Lambda_{\rm n}=\{e^{iq(z+h)}+e^{-iq(z+h)}\}$, $\theta$ is the polar angle. Following the calculation procedure in progressive case, we can get the SSFR for plane standing second sound.
 \begin{align}
	F=&-\frac{2\pi^{\rm 3}a^{\rm 2}T_{\rm 1}^{\rm 2}}{9\beta_{\rm 0}^{\rm 4}(\hbar c)^{\rm 3}T_{\rm 0}^{\rm 2}}\sum_{n=0}^\infty4(-1)^{n+1}n(n+1)(n+2)\nonumber\\
	&\times \left[U_{\rm n}U_{\rm n+1}(qa)+V_{\rm n}(qa)V_{\rm n+1}(qa)\right]\sin(2qh)\nonumber\\
	&-\frac{\pi^{\rm 3}T_{\rm 1}^{\rm 2}}{5\beta_{\rm 0}^{\rm 4}(\hbar c)^{\rm 3}q^{\rm 2}T_{\rm 0}^{\rm 2}}\sum_{n=0}^\infty 4(-1)^{n+1}(n+1)\nonumber\\
	&\times \left[U_{\rm n}U_{\rm n+1}(qa)+V_{\rm n}(qa)V_{\rm n+1}(qa)\right]\sin(2qh),
\end{align}
where $U$ and $V$ are determined by the scattering characteristics of the second sound at the boundary of the impurity particle. As we are examining the force effect on a stationary and adiabatic impurity particle, the expressions for $U$ and $V$ remain the same as in the progressive case. Consequently, the reduced SSRF can be expressed as follows:
 \begin{align}
	Y_{st}=&\frac{F}{AE_{\rm 1}^{\rm '}\sin(2qh)}=\frac{F}{\pi a^{\rm 2}E_{\rm 1}^{\rm '}\sin(2qh)}\nonumber\\
	&=-\frac{5T_{\rm 1}}{3T_{\rm 0}}\sum_{n=0}^\infty4(-1)^{n+1}n(n+1)(n+2)\nonumber\\
	&\times \left[U_{\rm n}U_{\rm n+1}(qa)+V_{\rm n}(qa)V_{\rm n+1}(qa)\right]\nonumber\\
	&-\frac{3T_{\rm 1}}{2T_{\rm 0}(qa)^{\rm 2}}\sum_{n=0}^\infty 4(-1)^{n+1}(n+1)\nonumber\\
	&\times \left[U_{\rm n}U_{\rm n+1}(qa)+V_{\rm n}(qa)V_{\rm n+1}(qa)\right],
\end{align}
where $A$ represents the cross-section area, and $E^{'}$ denotes the characteristic energy density of the incident second sound~\cite{tan2023tunable}. The results of $Y_{st}$ for different values of $T_1/T_0$ are illustrated in Figure~\ref{img11}(a). Unlike the progressive case~\cite{marston2006axial,2008Negative,mitri2009langevin,Marston2009Radiation,zhang2011geometrical,gong2021non,fan2021phase}, where only positive $Y_{p}$ values are observed, here, even in the case of plane standing second sound, negative $Y_{st}$ values can be obtained. In the context of standing second sound, where there is no specific propagation direction, a negative $Y_{st}$ value indicates that the direction of SSRF is towards the temperature antinode. Conversely, a positive $Y_{st}$ value signifies that the force direction is towards the temperature node. This stands in contrast to the progressive case. It is worth noting that, in the presence of a plane standing sound wave field, the reduced SSRF can exhibit both positive and negative values~\cite{hasegawa1979acoustic}, further supporting the validity of our findings.

Next, when the incident field is a standing zeroth-order Bessel second sound, the temperature field can be expressed as:
$T_{\rm inc}=T_1e^{-i\omega t}\left\{e^{iq_z(z+h)}+e^{-iq_z(z+h)}\right\}J_0(q_r\sin\theta)$.
Here, $q_z$ and $q_r$ represent the axial and radial wave numbers, respectively, i.e., $q=\sqrt{q_z^2+q_r^2}$. $J_0(x)$ denotes the cylindrical Bessel function of zeroth order. Similarly, in spherical coordinates, we obtain
\begin{align}
  T_{\rm inc}=T_{\rm 1}\sum_{n=0}^{\rm \infty}(2n+1)\Lambda_{\rm n}i^{\rm n}j_{\rm n}(qr)P_{\rm n}(\cos\theta)e^{-i\omega t},
\end{align}
where $j_n(x)$ is the spherical Bessel function of order $\rm n$, $P_{\rm n}(x)$ are the Legendre polynomial of order $\rm n$. Here, $\Lambda_{\rm n}=\{e^{iq_z(z+h)}+e^{-iq_z(z+h)}\}P_{\rm n}(\cos\delta)$, $\delta=\cos^{-1}(q_z/q)$ is the half cone angle of incident second sound. It is easy to verify that 
\begin{align}
	Y_{st}&=-\frac{5T_{\rm 1}}{3T_{\rm 0}}\sum_{n=0}^\infty4(-1)^{n+1}(n+1)\left[U_{\rm n}U_{\rm n+1}(qa)\right.\nonumber\\
	&\left.+V_{\rm n}(qa)V_{\rm n+1}(qa)\right]P_{\rm n}\cos(\delta)P_{\rm n+1}\cos(\delta)\nonumber\\
	&-\frac{3T_{\rm 1}}{2T_{\rm 0}(qa)^{\rm 2}}\sum_{n=0}^\infty 4(-1)^{n+1}n(n+1)(n+2)\left[U_{\rm n}U_{\rm n+1}(qa)\right.\nonumber\\
	&\left.+V_{\rm n}(qa)V_{\rm n+1}(qa)\right]P_{\rm n}\cos(\delta)P_{\rm n+1}\cos(\delta) ,
\end{align}
Here, the standing temperature field can be seen as the superposition of two progressive second sounds with the same half cone angle $\delta$. The variation of $Y_{st}$ with $qa$ is shown in Fig.~\ref{img11}(b). In contrast to the progressive zeroth-order Bessel second sound, $Y_{st}$ can be negative in this case, indicating the same behavior as in the plane standing case. Meanwhile, when $\delta=0$, the result is equivalent to the plane standing case, as the zeroth-order Bessel second sound reduces to plane standing second sound. The comparison with the acoustic radiation force under a zeroth-order Bessel standing sound wave field confirms the reasonableness of our results~\cite{mitri2008acoustic}.

\begin{figure*}[!ht]
	\includegraphics[width=1.0\linewidth]{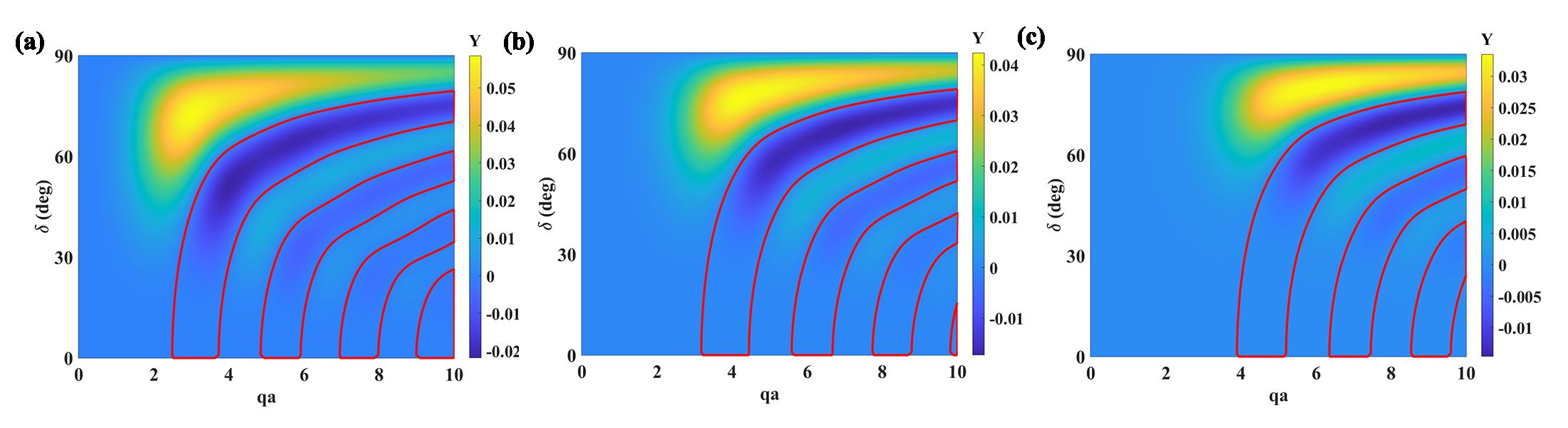}
	\caption{Numerical results of the reduced second sound radiation force $Y_{st}$ under the conditions of (a) second-order Bessel standing second sound as a function of $qa$ and $\delta$, (b) third-order Bessel standing second sound as a function of $qa$ and $\delta$, and (c) fourth-order Bessel standing second sound as a function of $qa$ and $\delta$, with $T_{\rm 1}/T_{\rm 0}=0.10$. The area enclosed by the red line is where $Y_{st}<0$.}
	\label{img2}
\end{figure*}
Further, we investigate the SSRF for first-order Bessel standing second sound. The incident second sound can be mathematically represented as $T_{\rm inc}=T_1e^{-i\omega t}\{e^{iq_z(z+h)}+e^{-iq_z(z+h)}\}J_1(q_r\sin\theta)e^{i\phi}$, where $\phi$ denotes the azimuth angle. By transforming this expression into spherical coordinates, we can derive the following results.
\begin{align}
  T_{\rm inc}&=T_{\rm 1}\sum_{n=1}^{\rm \infty}\frac{(n-1)!}{(n+1)!}(2n+1)\Lambda_{\rm n}i^{\rm n-1}j_{\rm n}(qr)\nonumber\\
  &\times P_{\rm n}^1(\cos\theta)e^{-i\omega t}e^{i\phi}.
\end{align}
Here, $\Lambda_{\rm n}=\{e^{iq_z(z+h)}+(-1)^{n+1}e^{-iq_z(z+h)}\}P_{\rm n}^1(\cos\delta)$. Following the same procedure as before, the reduced SSRF is
\begin{align}
	Y_{st}&=-\frac{5T_{\rm 1}}{3T_{\rm 0}}\sum_{n=1}^\infty \frac{4(-1)^{n+2}}{n+1}\left[U_{\rm n}U_{\rm n+1}(qa)\right.\nonumber\\
	&\left.+V_{\rm n}(qa)V_{\rm n+1}(qa)\right]P_{\rm n}\cos(\delta)P_{\rm n+1}\cos(\delta)\nonumber\\
	&-\frac{3T_{\rm 1}}{2T_{\rm 0}(qa)^{\rm 2}}\sum_{n=1}^\infty \frac{4(-1)^{n+2}n(n+2)}{n+1}\left[U_{\rm n}U_{\rm n+1}(qa)\right.\nonumber\\
	&\left.+V_{\rm n}(qa)V_{\rm n+1}(qa)\right]P_{\rm n}\cos(\delta)P_{\rm n+1}\cos(\delta) .
\end{align}
The results demonstrate that $Y_{st}$ can exhibit both positive and negative values. This finding aligns with the acoustic radiation force observed under a one-order Bessel standing sound wave field, as reported by Mitri et al.~\cite{mitri2009acoustic}.

As previously mentioned, the standing second sound can be conceptualized as the superposition of two second sounds propagating with the same amplitude. When these two second sounds possess different amplitudes, they give rise to a quasi-standing temperature field. For instance, considering the case of a plane quasi-standing field, the incident temperature field can be mathematically represented as $T_{\rm inc}=e^{-i\omega t}\{T_1e^{iq(z+h)}+T_2e^{-iq(z+h)}\}$, where we assume $T_1>T_2$. By employing the direct calculation method, we can determine the SSRF for the plane quasi-standing temperature field.
\begin{align}
	Y_{qst}&=-\left(\frac{T_2\sin(2qh)}{T_1}\right)\frac{5T_{\rm 1}}{3T_{\rm 0}}\sum_{n=0}^\infty4(-1)^{n+1}n(n+1)(n+2)\nonumber\\
	&\times \left[U_{\rm n}U_{\rm n+1}(qa)+V_{\rm n}(qa)V_{\rm n+1}(qa)\right]\nonumber\\
	&-(\frac{T_2\sin(2qh}{T_1})\frac{3T_{\rm 1}}{2T_{\rm 0}(qa)^{\rm 2}}\sum_{n=0}^\infty 4(-1)^{n+1}(n+1)\nonumber\\
	&\times \left[U_{\rm n}U_{\rm n+1}(qa)+V_{\rm n}(qa)V_{\rm n+1}(qa)\right]\nonumber\\
  &-\left(1-\frac{T_2^2}{T_1^2}\right)\frac{5T_{\rm 1}}{3T_{\rm 0}}\sum_{n=0}^\infty2(n+1)\nonumber\\
	&\times \left[V_{\rm n}(qa)U_{\rm n+1}(qa)-U_{\rm n}(qa)V_{\rm n+1}(qa)\right]\nonumber\\
	&-\left(1-\frac{T_2^2}{T_1^2}\right)\frac{3T_{\rm 1}}{2T_{\rm 0}(qa)^{\rm 2}}\sum_{n=0}^\infty 2n(n+1)(n+2)\nonumber\\
	&\times \left[V_{\rm n}(qa)U_{\rm n+1}(qa)-U_{\rm n}(qa)V_{\rm n+1}(qa)\right],
\end{align}
Combining the findings from the progressive and standing cases, we observe that $Y_{qst}=Y_p(1-\frac{T_2^2}{T_1^2})+Y_{st}(\frac{T_2\sin(2qh)}{T_1})$. In this equation, $Y_p$ represents the reduced SSRF under plane progressive second sound. This expression serves as a link between the results obtained from the progressive and standing cases. Importantly, it is applicable not only to the plane case but also to the Bessel temperature field. The only difference is that $\sin(2qh)$ should be replaced by $\sin(2q_zh)$. We see when $T_1=T_2$, the result coincides with the standing case, as the quasi-standing temperature field reduces to a standing temperature field. Furthermore, if $T_1=0$, then $Y_{st}$ is equivalent to the SSRF for progressive second sound.

\section{Discussion and conclusion}

In this paper, we have conducted a study on the SSRF (Second Sound Radiation Force) under the standing and quasi-standing temperature field. In the case of a standing second sound field, the reduced SSRF can have both positive and negative values, even for the plane and zeroth-order Bessel standing temperature field. Similar results are found under high-order Bessel standing cases, as shown in Fig.~\ref{img2}(a)-Fig.~\ref{img2}(c). However, it is important to note that the sign of $Y_{st}$ indicates the direction relative to the temperature antinode or node, which is distinct from the progressive cases. We have also demonstrated that the results of the quasi-standing temperature field can link the progressive case and standing case, which is of theoretical significance considering the existence of this relation even for high-order Bessel temperature field.

Nevertheless, our focus in this paper remains on the hydrodynamic regime of thermal conduction force. For further studies, we could extend our analysis to other regimes, such as the diffusive and casir regimes~\cite{sorbello1972phonon,gerl1971calculation}. Additionally, recent research has shown that density variation can induce a force effect for the fluid~\cite{karlsen2016acoustic}. In the case of our solid-solid system, where we assume the background temperature is constant, we have neglected the corresponding density variation and have not considered the force effect. Our primary focus is on the interaction between second sound and impurity. Furthermore, we have assumed that the impurity is fixed, and there is a need for exploration regarding the force effect on movable impurities. Moreover, studying the force effect on non-adiabatic impurity particles proves to be an interesting research topic~\cite{gao2007magnetophoresis,xu2018thermal1,dong2004dielectric,huang2003dielectrophoresis,ye2008non,shen2016thermal1}.

In summary, building upon the research on second sound radiation force for progressive second sound, we have extended our analysis to incorporate the standing and quasi-standing cases. We have revealed distinct characteristics of the second radiation force for these cases. Additionally, we have explored the relationship between the progressive, standing, and quasi-standing cases. This not only enriches the fundamental theory of second sound radiation force but also paves the way for future studies in more complex scenarios~\cite{liu2013statistical,huang2005magneto,qiu2015nonstraight}. In terms of applications, our study offers people an additional degree of freedom to manipulate the local thermal conductivity of dielectric crystals.
\begin{acknowledgments}
  We acknowledge the financial support provided by the National Natural Science Foundation of China under Grants No. 12035004 and No. 12320101004, the Science and Technology Commission of Shanghai Municipality under Grant No. 20JC1414700, and the Innovation Program of Shanghai Municipal Education Commission under Grant No. 2023ZKZD06.
\end{acknowledgments}



\begin{thebibliography}{71}%
\makeatletter
\providecommand \@ifxundefined [1]{%
 \@ifx{#1\undefined}
}%
\providecommand \@ifnum [1]{%
 \ifnum #1\expandafter \@firstoftwo
 \else \expandafter \@secondoftwo
 \fi
}%
\providecommand \@ifx [1]{%
 \ifx #1\expandafter \@firstoftwo
 \else \expandafter \@secondoftwo
 \fi
}%
\providecommand \natexlab [1]{#1}%
\providecommand \enquote  [1]{``#1''}%
\providecommand \bibnamefont  [1]{#1}%
\providecommand \bibfnamefont [1]{#1}%
\providecommand \citenamefont [1]{#1}%
\providecommand \href@noop [0]{\@secondoftwo}%
\providecommand \href [0]{\begingroup \@sanitize@url \@href}%
\providecommand \@href[1]{\@@startlink{#1}\@@href}%
\providecommand \@@href[1]{\endgroup#1\@@endlink}%
\providecommand \@sanitize@url [0]{\catcode `\\12\catcode `\$12\catcode
  `\&12\catcode `\#12\catcode `\^12\catcode `\_12\catcode `\%12\relax}%
\providecommand \@@startlink[1]{}%
\providecommand \@@endlink[0]{}%
\providecommand \url  [0]{\begingroup\@sanitize@url \@url }%
\providecommand \@url [1]{\endgroup\@href {#1}{\urlprefix }}%
\providecommand \urlprefix  [0]{URL }%
\providecommand \Eprint [0]{\href }%
\providecommand \doibase [0]{https://doi.org/}%
\providecommand \selectlanguage [0]{\@gobble}%
\providecommand \bibinfo  [0]{\@secondoftwo}%
\providecommand \bibfield  [0]{\@secondoftwo}%
\providecommand \translation [1]{[#1]}%
\providecommand \BibitemOpen [0]{}%
\providecommand \bibitemStop [0]{}%
\providecommand \bibitemNoStop [0]{.\EOS\space}%
\providecommand \EOS [0]{\spacefactor3000\relax}%
\providecommand \BibitemShut  [1]{\csname bibitem#1\endcsname}%
\let\auto@bib@innerbib\@empty
\bibitem [{\citenamefont {Zhou}\ \emph {et~al.}(2023)\citenamefont {Zhou},
  \citenamefont {Xu},\ and\ \citenamefont {Huang}}]{zhou2023adaptive}%
  \BibitemOpen
  \bibfield  {author} {\bibinfo {author} {\bibfnamefont {X.}~\bibnamefont
  {Zhou}}, \bibinfo {author} {\bibfnamefont {X.}~\bibnamefont {Xu}},\ and\
  \bibinfo {author} {\bibfnamefont {J.}~\bibnamefont {Huang}},\ }\bibfield
  {title} {\bibinfo {title} {Adaptive multi-temperature control for transport
  and storage containers enabled by phase-change materials},\ }\href@noop {}
  {\bibfield  {journal} {\bibinfo  {journal} {Nat. Commun.}\ }\textbf {\bibinfo
  {volume} {14}},\ \bibinfo {pages} {5449} (\bibinfo {year}
  {2023})}\BibitemShut {NoStop}%
\bibitem [{\citenamefont {Xu}\ \emph {et~al.}(2023)\citenamefont {Xu},
  \citenamefont {Liu}, \citenamefont {Xu}, \citenamefont {Huang},\ and\
  \citenamefont {Qiu}}]{xu2023giant}%
  \BibitemOpen
  \bibfield  {author} {\bibinfo {author} {\bibfnamefont {L.}~\bibnamefont
  {Xu}}, \bibinfo {author} {\bibfnamefont {J.}~\bibnamefont {Liu}}, \bibinfo
  {author} {\bibfnamefont {G.}~\bibnamefont {Xu}}, \bibinfo {author}
  {\bibfnamefont {J.}~\bibnamefont {Huang}},\ and\ \bibinfo {author}
  {\bibfnamefont {C.-W.}\ \bibnamefont {Qiu}},\ }\bibfield  {title} {\bibinfo
  {title} {Giant, magnet-free, and room-temperature hall-like heat transfer},\
  }\href@noop {} {\bibfield  {journal} {\bibinfo  {journal} {Proc. Natl. Acad.
  Sci. USA}\ }\textbf {\bibinfo {volume} {120}},\ \bibinfo {pages}
  {e2305755120} (\bibinfo {year} {2023})}\BibitemShut {NoStop}%
\bibitem [{\citenamefont {Yang}\ \emph {et~al.}(2023)\citenamefont {Yang},
  \citenamefont {Zhang}, \citenamefont {Xu}, \citenamefont {Liu}, \citenamefont
  {Jin}, \citenamefont {Zhuang}, \citenamefont {Lei}, \citenamefont {Liu},
  \citenamefont {Jiang}, \citenamefont {Ouyang} \emph
  {et~al.}}]{yang2023controlling}%
  \BibitemOpen
  \bibfield  {author} {\bibinfo {author} {\bibfnamefont {F.}~\bibnamefont
  {Yang}}, \bibinfo {author} {\bibfnamefont {Z.}~\bibnamefont {Zhang}},
  \bibinfo {author} {\bibfnamefont {L.}~\bibnamefont {Xu}}, \bibinfo {author}
  {\bibfnamefont {Z.}~\bibnamefont {Liu}}, \bibinfo {author} {\bibfnamefont
  {P.}~\bibnamefont {Jin}}, \bibinfo {author} {\bibfnamefont {P.}~\bibnamefont
  {Zhuang}}, \bibinfo {author} {\bibfnamefont {M.}~\bibnamefont {Lei}},
  \bibinfo {author} {\bibfnamefont {J.}~\bibnamefont {Liu}}, \bibinfo {author}
  {\bibfnamefont {J.-H.}\ \bibnamefont {Jiang}}, \bibinfo {author}
  {\bibfnamefont {X.}~\bibnamefont {Ouyang}}, \emph {et~al.},\ }\bibfield
  {title} {\bibinfo {title} {Controlling mass and energy diffusion with
  metamaterials},\ }\href@noop {} {\bibfield  {journal} {\bibinfo  {journal}
  {arXiv preprint arXiv:2309.04711}\ } (\bibinfo {year} {2023})}\BibitemShut
  {NoStop}%
\bibitem [{\citenamefont {Xu}\ \emph {et~al.}(2022{\natexlab{a}})\citenamefont
  {Xu}, \citenamefont {Xu}, \citenamefont {Huang},\ and\ \citenamefont
  {Qiu}}]{xu2022diffusive}%
  \BibitemOpen
  \bibfield  {author} {\bibinfo {author} {\bibfnamefont {L.}~\bibnamefont
  {Xu}}, \bibinfo {author} {\bibfnamefont {G.}~\bibnamefont {Xu}}, \bibinfo
  {author} {\bibfnamefont {J.}~\bibnamefont {Huang}},\ and\ \bibinfo {author}
  {\bibfnamefont {C.-W.}\ \bibnamefont {Qiu}},\ }\bibfield  {title} {\bibinfo
  {title} {Diffusive fizeau drag in spatiotemporal thermal metamaterials},\
  }\href@noop {} {\bibfield  {journal} {\bibinfo  {journal} {Phys. Rev. Lett.}\
  }\textbf {\bibinfo {volume} {128}},\ \bibinfo {pages} {145901} (\bibinfo
  {year} {2022}{\natexlab{a}})}\BibitemShut {NoStop}%
\bibitem [{\citenamefont {Xu}\ \emph {et~al.}(2022{\natexlab{b}})\citenamefont
  {Xu}, \citenamefont {Xu}, \citenamefont {Li}, \citenamefont {Li},
  \citenamefont {Huang},\ and\ \citenamefont {Qiu}}]{xu2022thermal}%
  \BibitemOpen
  \bibfield  {author} {\bibinfo {author} {\bibfnamefont {L.}~\bibnamefont
  {Xu}}, \bibinfo {author} {\bibfnamefont {G.}~\bibnamefont {Xu}}, \bibinfo
  {author} {\bibfnamefont {J.}~\bibnamefont {Li}}, \bibinfo {author}
  {\bibfnamefont {Y.}~\bibnamefont {Li}}, \bibinfo {author} {\bibfnamefont
  {J.}~\bibnamefont {Huang}},\ and\ \bibinfo {author} {\bibfnamefont {C.-W.}\
  \bibnamefont {Qiu}},\ }\bibfield  {title} {\bibinfo {title} {Thermal willis
  coupling in spatiotemporal diffusive metamaterials},\ }\href@noop {}
  {\bibfield  {journal} {\bibinfo  {journal} {Phys. Rev. Lett.}\ }\textbf
  {\bibinfo {volume} {129}},\ \bibinfo {pages} {155901} (\bibinfo {year}
  {2022}{\natexlab{b}})}\BibitemShut {NoStop}%
\bibitem [{\citenamefont {Jin}\ \emph {et~al.}(2023)\citenamefont {Jin},
  \citenamefont {Liu}, \citenamefont {Xu}, \citenamefont {Wang}, \citenamefont
  {Ouyang}, \citenamefont {Jiang},\ and\ \citenamefont
  {Huang}}]{jin2023tunable}%
  \BibitemOpen
  \bibfield  {author} {\bibinfo {author} {\bibfnamefont {P.}~\bibnamefont
  {Jin}}, \bibinfo {author} {\bibfnamefont {J.}~\bibnamefont {Liu}}, \bibinfo
  {author} {\bibfnamefont {L.}~\bibnamefont {Xu}}, \bibinfo {author}
  {\bibfnamefont {J.}~\bibnamefont {Wang}}, \bibinfo {author} {\bibfnamefont
  {X.}~\bibnamefont {Ouyang}}, \bibinfo {author} {\bibfnamefont {J.-H.}\
  \bibnamefont {Jiang}},\ and\ \bibinfo {author} {\bibfnamefont
  {J.}~\bibnamefont {Huang}},\ }\bibfield  {title} {\bibinfo {title} {Tunable
  liquid--solid hybrid thermal metamaterials with a topology transition},\
  }\href@noop {} {\bibfield  {journal} {\bibinfo  {journal} {Proc. Natl. Acad.
  Sci. USA}\ }\textbf {\bibinfo {volume} {120}},\ \bibinfo {pages}
  {e2217068120} (\bibinfo {year} {2023})}\BibitemShut {NoStop}%
\bibitem [{\citenamefont {Zhang}\ \emph {et~al.}(2023)\citenamefont {Zhang},
  \citenamefont {Xu}, \citenamefont {Qu}, \citenamefont {Lei}, \citenamefont
  {Lin}, \citenamefont {Ouyang}, \citenamefont {Jiang},\ and\ \citenamefont
  {Huang}}]{zhang2023diffusion}%
  \BibitemOpen
  \bibfield  {author} {\bibinfo {author} {\bibfnamefont {Z.}~\bibnamefont
  {Zhang}}, \bibinfo {author} {\bibfnamefont {L.}~\bibnamefont {Xu}}, \bibinfo
  {author} {\bibfnamefont {T.}~\bibnamefont {Qu}}, \bibinfo {author}
  {\bibfnamefont {M.}~\bibnamefont {Lei}}, \bibinfo {author} {\bibfnamefont
  {Z.-K.}\ \bibnamefont {Lin}}, \bibinfo {author} {\bibfnamefont
  {X.}~\bibnamefont {Ouyang}}, \bibinfo {author} {\bibfnamefont {J.-H.}\
  \bibnamefont {Jiang}},\ and\ \bibinfo {author} {\bibfnamefont
  {J.}~\bibnamefont {Huang}},\ }\bibfield  {title} {\bibinfo {title} {Diffusion
  metamaterials},\ }\href@noop {} {\bibfield  {journal} {\bibinfo  {journal}
  {Nat. Rev. Phys.}\ }\textbf {\bibinfo {volume} {5}},\ \bibinfo {pages} {218}
  (\bibinfo {year} {2023})}\BibitemShut {NoStop}%
\bibitem [{\citenamefont {Li}\ \emph {et~al.}(2021)\citenamefont {Li},
  \citenamefont {Li}, \citenamefont {Han}, \citenamefont {Zheng}, \citenamefont
  {Li}, \citenamefont {Li}, \citenamefont {Fan},\ and\ \citenamefont
  {Qiu}}]{li2021transforming}%
  \BibitemOpen
  \bibfield  {author} {\bibinfo {author} {\bibfnamefont {Y.}~\bibnamefont
  {Li}}, \bibinfo {author} {\bibfnamefont {W.}~\bibnamefont {Li}}, \bibinfo
  {author} {\bibfnamefont {T.}~\bibnamefont {Han}}, \bibinfo {author}
  {\bibfnamefont {X.}~\bibnamefont {Zheng}}, \bibinfo {author} {\bibfnamefont
  {J.}~\bibnamefont {Li}}, \bibinfo {author} {\bibfnamefont {B.}~\bibnamefont
  {Li}}, \bibinfo {author} {\bibfnamefont {S.}~\bibnamefont {Fan}},\ and\
  \bibinfo {author} {\bibfnamefont {C.-W.}\ \bibnamefont {Qiu}},\ }\bibfield
  {title} {\bibinfo {title} {Transforming heat transfer with thermal
  metamaterials and devices},\ }\href@noop {} {\bibfield  {journal} {\bibinfo
  {journal} {Nat. Rev. Mater.}\ }\textbf {\bibinfo {volume} {6}},\ \bibinfo
  {pages} {488} (\bibinfo {year} {2021})}\BibitemShut {NoStop}%
\bibitem [{\citenamefont {Palla}\ \emph {et~al.}(2020)\citenamefont {Palla},
  \citenamefont {Patera}, \citenamefont {Cleri},\ and\ \citenamefont
  {Giordano}}]{palla2020stochastic}%
  \BibitemOpen
  \bibfield  {author} {\bibinfo {author} {\bibfnamefont {P.~L.}\ \bibnamefont
  {Palla}}, \bibinfo {author} {\bibfnamefont {G.}~\bibnamefont {Patera}},
  \bibinfo {author} {\bibfnamefont {F.}~\bibnamefont {Cleri}},\ and\ \bibinfo
  {author} {\bibfnamefont {S.}~\bibnamefont {Giordano}},\ }\bibfield  {title}
  {\bibinfo {title} {A stochastic force model for the ballistic-diffusive
  transition of heat conduction},\ }\href@noop {} {\bibfield  {journal}
  {\bibinfo  {journal} {Physica Scripta}\ }\textbf {\bibinfo {volume} {95}},\
  \bibinfo {pages} {075703} (\bibinfo {year} {2020})}\BibitemShut {NoStop}%
\bibitem [{\citenamefont {Lepri}\ \emph {et~al.}(2003)\citenamefont {Lepri},
  \citenamefont {Livi},\ and\ \citenamefont {Politi}}]{lepri2003thermal}%
  \BibitemOpen
  \bibfield  {author} {\bibinfo {author} {\bibfnamefont {S.}~\bibnamefont
  {Lepri}}, \bibinfo {author} {\bibfnamefont {R.}~\bibnamefont {Livi}},\ and\
  \bibinfo {author} {\bibfnamefont {A.}~\bibnamefont {Politi}},\ }\bibfield
  {title} {\bibinfo {title} {Thermal conduction in classical low-dimensional
  lattices},\ }\href@noop {} {\bibfield  {journal} {\bibinfo  {journal} {Phys.
  Rep.}\ }\textbf {\bibinfo {volume} {377}},\ \bibinfo {pages} {1} (\bibinfo
  {year} {2003})}\BibitemShut {NoStop}%
\bibitem [{\citenamefont {Dhar}(2008)}]{dhar2008heat}%
  \BibitemOpen
  \bibfield  {author} {\bibinfo {author} {\bibfnamefont {A.}~\bibnamefont
  {Dhar}},\ }\bibfield  {title} {\bibinfo {title} {Heat transport in
  low-dimensional systems},\ }\href@noop {} {\bibfield  {journal} {\bibinfo
  {journal} {Adv. Phys.}\ }\textbf {\bibinfo {volume} {57}},\ \bibinfo {pages}
  {457} (\bibinfo {year} {2008})}\BibitemShut {NoStop}%
\bibitem [{\citenamefont {Yang}\ \emph {et~al.}(2017)\citenamefont {Yang},
  \citenamefont {Xu}, \citenamefont {Wang},\ and\ \citenamefont
  {Huang}}]{yang2017full}%
  \BibitemOpen
  \bibfield  {author} {\bibinfo {author} {\bibfnamefont {S.}~\bibnamefont
  {Yang}}, \bibinfo {author} {\bibfnamefont {L.}~\bibnamefont {Xu}}, \bibinfo
  {author} {\bibfnamefont {R.}~\bibnamefont {Wang}},\ and\ \bibinfo {author}
  {\bibfnamefont {J.}~\bibnamefont {Huang}},\ }\bibfield  {title} {\bibinfo
  {title} {Full control of heat transfer in single-particle structural
  materials},\ }\href@noop {} {\bibfield  {journal} {\bibinfo  {journal} {Appl.
  Phys. Lett.}\ }\textbf {\bibinfo {volume} {111}} (\bibinfo {year}
  {2017})}\BibitemShut {NoStop}%
\bibitem [{\citenamefont {Xu}\ and\ \citenamefont
  {Huang}(2020)}]{xu2020active}%
  \BibitemOpen
  \bibfield  {author} {\bibinfo {author} {\bibfnamefont {L.-J.}\ \bibnamefont
  {Xu}}\ and\ \bibinfo {author} {\bibfnamefont {J.-P.}\ \bibnamefont {Huang}},\
  }\bibfield  {title} {\bibinfo {title} {Active thermal wave cloak},\
  }\href@noop {} {\bibfield  {journal} {\bibinfo  {journal} {Chin. Phys.
  Lett.}\ }\textbf {\bibinfo {volume} {37}},\ \bibinfo {pages} {120501}
  (\bibinfo {year} {2020})}\BibitemShut {NoStop}%
\bibitem [{\citenamefont {Xu}\ \emph {et~al.}(2021)\citenamefont {Xu},
  \citenamefont {Wang}, \citenamefont {Dai}, \citenamefont {Yang},
  \citenamefont {Yang}, \citenamefont {Wang},\ and\ \citenamefont
  {Huang}}]{xu2021geometric}%
  \BibitemOpen
  \bibfield  {author} {\bibinfo {author} {\bibfnamefont {L.}~\bibnamefont
  {Xu}}, \bibinfo {author} {\bibfnamefont {J.}~\bibnamefont {Wang}}, \bibinfo
  {author} {\bibfnamefont {G.}~\bibnamefont {Dai}}, \bibinfo {author}
  {\bibfnamefont {S.}~\bibnamefont {Yang}}, \bibinfo {author} {\bibfnamefont
  {F.}~\bibnamefont {Yang}}, \bibinfo {author} {\bibfnamefont {G.}~\bibnamefont
  {Wang}},\ and\ \bibinfo {author} {\bibfnamefont {J.}~\bibnamefont {Huang}},\
  }\bibfield  {title} {\bibinfo {title} {Geometric phase, effective
  conductivity enhancement, and invisibility cloak in thermal
  convection-conduction},\ }\href@noop {} {\bibfield  {journal} {\bibinfo
  {journal} {Int. J. Heat Mass Transfer}\ }\textbf {\bibinfo {volume} {165}},\
  \bibinfo {pages} {120659} (\bibinfo {year} {2021})}\BibitemShut {NoStop}%
\bibitem [{\citenamefont {Xu}\ \emph {et~al.}(2020{\natexlab{a}})\citenamefont
  {Xu}, \citenamefont {Yang}, \citenamefont {Dai},\ and\ \citenamefont
  {Huang}}]{xu2020transformation}%
  \BibitemOpen
  \bibfield  {author} {\bibinfo {author} {\bibfnamefont {L.}~\bibnamefont
  {Xu}}, \bibinfo {author} {\bibfnamefont {S.}~\bibnamefont {Yang}}, \bibinfo
  {author} {\bibfnamefont {G.}~\bibnamefont {Dai}},\ and\ \bibinfo {author}
  {\bibfnamefont {J.}~\bibnamefont {Huang}},\ }\bibfield  {title} {\bibinfo
  {title} {Transformation omnithermotics: simultaneous manipulation of three
  basic modes of heat transfer},\ }\href@noop {} {\bibfield  {journal}
  {\bibinfo  {journal} {ES Energy \& Environment}\ }\textbf {\bibinfo {volume}
  {7}},\ \bibinfo {pages} {65} (\bibinfo {year}
  {2020}{\natexlab{a}})}\BibitemShut {NoStop}%
\bibitem [{\citenamefont {Xu}\ \emph {et~al.}(2020{\natexlab{b}})\citenamefont
  {Xu}, \citenamefont {Dai},\ and\ \citenamefont
  {Huang}}]{xu2020transformation1}%
  \BibitemOpen
  \bibfield  {author} {\bibinfo {author} {\bibfnamefont {L.}~\bibnamefont
  {Xu}}, \bibinfo {author} {\bibfnamefont {G.}~\bibnamefont {Dai}},\ and\
  \bibinfo {author} {\bibfnamefont {J.}~\bibnamefont {Huang}},\ }\bibfield
  {title} {\bibinfo {title} {Transformation multithermotics: controlling
  radiation and conduction simultaneously},\ }\href@noop {} {\bibfield
  {journal} {\bibinfo  {journal} {Phys. Rev. Appl.}\ }\textbf {\bibinfo
  {volume} {13}},\ \bibinfo {pages} {024063} (\bibinfo {year}
  {2020}{\natexlab{b}})}\BibitemShut {NoStop}%
\bibitem [{\citenamefont {Dai}\ and\ \citenamefont
  {Huang}(2018)}]{dai2018transient}%
  \BibitemOpen
  \bibfield  {author} {\bibinfo {author} {\bibfnamefont {G.}~\bibnamefont
  {Dai}}\ and\ \bibinfo {author} {\bibfnamefont {J.}~\bibnamefont {Huang}},\
  }\bibfield  {title} {\bibinfo {title} {A transient regime for transforming
  thermal convection: Cloaking, concentrating, and rotating creeping flow and
  heat flux},\ }\href@noop {} {\bibfield  {journal} {\bibinfo  {journal} {J.
  Appl. Phys.}\ }\textbf {\bibinfo {volume} {124}} (\bibinfo {year}
  {2018})}\BibitemShut {NoStop}%
\bibitem [{\citenamefont {Zhu}\ \emph {et~al.}(2015)\citenamefont {Zhu},
  \citenamefont {Shen},\ and\ \citenamefont {Huang}}]{zhu2015converting}%
  \BibitemOpen
  \bibfield  {author} {\bibinfo {author} {\bibfnamefont {N.}~\bibnamefont
  {Zhu}}, \bibinfo {author} {\bibfnamefont {X.}~\bibnamefont {Shen}},\ and\
  \bibinfo {author} {\bibfnamefont {J.}~\bibnamefont {Huang}},\ }\bibfield
  {title} {\bibinfo {title} {Converting the patterns of local heat flux via
  thermal illusion device},\ }\href@noop {} {\bibfield  {journal} {\bibinfo
  {journal} {AIP Adv.}\ }\textbf {\bibinfo {volume} {5}} (\bibinfo {year}
  {2015})}\BibitemShut {NoStop}%
\bibitem [{\citenamefont {Yang}\ \emph {et~al.}(2019)\citenamefont {Yang},
  \citenamefont {Xu},\ and\ \citenamefont {Huang}}]{yang2019thermal}%
  \BibitemOpen
  \bibfield  {author} {\bibinfo {author} {\bibfnamefont {F.}~\bibnamefont
  {Yang}}, \bibinfo {author} {\bibfnamefont {L.}~\bibnamefont {Xu}},\ and\
  \bibinfo {author} {\bibfnamefont {J.}~\bibnamefont {Huang}},\ }\bibfield
  {title} {\bibinfo {title} {Thermal illusion of porous media with
  convection-diffusion process: transparency, concentrating, and cloaking},\
  }\href@noop {} {\bibfield  {journal} {\bibinfo  {journal} {ES Energy \&
  Environment}\ }\textbf {\bibinfo {volume} {6}},\ \bibinfo {pages} {45}
  (\bibinfo {year} {2019})}\BibitemShut {NoStop}%
\bibitem [{\citenamefont {Xu}\ \emph {et~al.}(2019)\citenamefont {Xu},
  \citenamefont {Yang},\ and\ \citenamefont {Huang}}]{xu2019thermal}%
  \BibitemOpen
  \bibfield  {author} {\bibinfo {author} {\bibfnamefont {L.}~\bibnamefont
  {Xu}}, \bibinfo {author} {\bibfnamefont {S.}~\bibnamefont {Yang}},\ and\
  \bibinfo {author} {\bibfnamefont {J.}~\bibnamefont {Huang}},\ }\bibfield
  {title} {\bibinfo {title} {Thermal transparency induced by periodic
  interparticle interaction},\ }\href@noop {} {\bibfield  {journal} {\bibinfo
  {journal} {Phys. Rev. Appl.}\ }\textbf {\bibinfo {volume} {11}},\ \bibinfo
  {pages} {034056} (\bibinfo {year} {2019})}\BibitemShut {NoStop}%
\bibitem [{\citenamefont {Jin}\ \emph {et~al.}(2020)\citenamefont {Jin},
  \citenamefont {Xu}, \citenamefont {Jiang}, \citenamefont {Zhang},\ and\
  \citenamefont {Huang}}]{jin2020making}%
  \BibitemOpen
  \bibfield  {author} {\bibinfo {author} {\bibfnamefont {P.}~\bibnamefont
  {Jin}}, \bibinfo {author} {\bibfnamefont {L.}~\bibnamefont {Xu}}, \bibinfo
  {author} {\bibfnamefont {T.}~\bibnamefont {Jiang}}, \bibinfo {author}
  {\bibfnamefont {L.}~\bibnamefont {Zhang}},\ and\ \bibinfo {author}
  {\bibfnamefont {J.}~\bibnamefont {Huang}},\ }\bibfield  {title} {\bibinfo
  {title} {Making thermal sensors accurate and invisible with an anisotropic
  monolayer scheme},\ }\href@noop {} {\bibfield  {journal} {\bibinfo  {journal}
  {Int. J. Heat Mass Transfer}\ }\textbf {\bibinfo {volume} {163}},\ \bibinfo
  {pages} {120437} (\bibinfo {year} {2020})}\BibitemShut {NoStop}%
\bibitem [{\citenamefont {Gaeta}(1969)}]{gaeta1969radiation}%
  \BibitemOpen
  \bibfield  {author} {\bibinfo {author} {\bibfnamefont {F.}~\bibnamefont
  {Gaeta}},\ }\bibfield  {title} {\bibinfo {title} {Radiation pressure theory
  of thermal diffusion in liquids},\ }\href@noop {} {\bibfield  {journal}
  {\bibinfo  {journal} {Phys. Rev.}\ }\textbf {\bibinfo {volume} {182}},\
  \bibinfo {pages} {289} (\bibinfo {year} {1969})}\BibitemShut {NoStop}%
\bibitem [{\citenamefont {{Gaeta, F. S and Ascolese, E and Tomicki,
  B}}(1991)}]{gaeta1991radiation}%
  \BibitemOpen
  \bibfield  {author} {\bibinfo {author} {\bibnamefont {{Gaeta, F. S and
  Ascolese, E and Tomicki, B}}},\ }\bibfield  {title} {\bibinfo {title}
  {Radiation forces associated with heat propagation in nonisothermal
  systems},\ }\href@noop {} {\bibfield  {journal} {\bibinfo  {journal} {Phys.
  Rev. A}\ }\textbf {\bibinfo {volume} {44}},\ \bibinfo {pages} {5003}
  (\bibinfo {year} {1991})}\BibitemShut {NoStop}%
\bibitem [{\citenamefont {{Albanese, C and Dell'Aversana, P and Gaeta, F.
  S}}(1997)}]{albanese1997experimental}%
  \BibitemOpen
  \bibfield  {author} {\bibinfo {author} {\bibnamefont {{Albanese, C and
  Dell'Aversana, P and Gaeta, F. S}}},\ }\bibfield  {title} {\bibinfo {title}
  {Experimental detection of forces produced by the flow of heat},\ }\href@noop
  {} {\bibfield  {journal} {\bibinfo  {journal} {Phys. Rev. Lett.}\ }\textbf
  {\bibinfo {volume} {79}},\ \bibinfo {pages} {4151} (\bibinfo {year}
  {1997})}\BibitemShut {NoStop}%
\bibitem [{\citenamefont {Tan}\ \emph {et~al.}(2023)\citenamefont {Tan},
  \citenamefont {Qiu}, \citenamefont {Xu},\ and\ \citenamefont
  {Huang}}]{tan2023tunable}%
  \BibitemOpen
  \bibfield  {author} {\bibinfo {author} {\bibfnamefont {H.}~\bibnamefont
  {Tan}}, \bibinfo {author} {\bibfnamefont {Y.}~\bibnamefont {Qiu}}, \bibinfo
  {author} {\bibfnamefont {L.}~\bibnamefont {Xu}},\ and\ \bibinfo {author}
  {\bibfnamefont {J.}~\bibnamefont {Huang}},\ }\bibfield  {title} {\bibinfo
  {title} {Tunable thermal conduction force without macroscopic temperature
  gradients},\ }\href@noop {} {\bibfield  {journal} {\bibinfo  {journal} {Phys.
  Rev. E}\ }\textbf {\bibinfo {volume} {108}},\ \bibinfo {pages} {034105}
  (\bibinfo {year} {2023})}\BibitemShut {NoStop}%
\bibitem [{\citenamefont {Huberman}\ \emph {et~al.}(2019)\citenamefont
  {Huberman}, \citenamefont {Duncan}, \citenamefont {Chen}, \citenamefont
  {Song}, \citenamefont {Chiloyan}, \citenamefont {Ding}, \citenamefont
  {Maznev}, \citenamefont {Chen},\ and\ \citenamefont
  {Nelson}}]{huberman2019observation}%
  \BibitemOpen
  \bibfield  {author} {\bibinfo {author} {\bibfnamefont {S.}~\bibnamefont
  {Huberman}}, \bibinfo {author} {\bibfnamefont {R.~A.}\ \bibnamefont
  {Duncan}}, \bibinfo {author} {\bibfnamefont {K.}~\bibnamefont {Chen}},
  \bibinfo {author} {\bibfnamefont {B.}~\bibnamefont {Song}}, \bibinfo {author}
  {\bibfnamefont {V.}~\bibnamefont {Chiloyan}}, \bibinfo {author}
  {\bibfnamefont {Z.}~\bibnamefont {Ding}}, \bibinfo {author} {\bibfnamefont
  {A.~A.}\ \bibnamefont {Maznev}}, \bibinfo {author} {\bibfnamefont
  {G.}~\bibnamefont {Chen}},\ and\ \bibinfo {author} {\bibfnamefont {K.~A.}\
  \bibnamefont {Nelson}},\ }\bibfield  {title} {\bibinfo {title} {{Observation
  of second sound in graphite at temperatures above 100 K}},\ }\href@noop {}
  {\bibfield  {journal} {\bibinfo  {journal} {Science}\ }\textbf {\bibinfo
  {volume} {364}},\ \bibinfo {pages} {375} (\bibinfo {year}
  {2019})}\BibitemShut {NoStop}%
\bibitem [{\citenamefont {Ding}\ \emph {et~al.}(2022)\citenamefont {Ding},
  \citenamefont {Chen}, \citenamefont {Song}, \citenamefont {Shin},
  \citenamefont {Maznev}, \citenamefont {Nelson},\ and\ \citenamefont
  {Chen}}]{ding2022observation}%
  \BibitemOpen
  \bibfield  {author} {\bibinfo {author} {\bibfnamefont {Z.}~\bibnamefont
  {Ding}}, \bibinfo {author} {\bibfnamefont {K.}~\bibnamefont {Chen}}, \bibinfo
  {author} {\bibfnamefont {B.}~\bibnamefont {Song}}, \bibinfo {author}
  {\bibfnamefont {J.}~\bibnamefont {Shin}}, \bibinfo {author} {\bibfnamefont
  {A.~A.}\ \bibnamefont {Maznev}}, \bibinfo {author} {\bibfnamefont {K.~A.}\
  \bibnamefont {Nelson}},\ and\ \bibinfo {author} {\bibfnamefont
  {G.}~\bibnamefont {Chen}},\ }\bibfield  {title} {\bibinfo {title}
  {{Observation of second sound in graphite over 200 K}},\ }\href@noop {}
  {\bibfield  {journal} {\bibinfo  {journal} {Nat. Commun.}\ }\textbf {\bibinfo
  {volume} {13}},\ \bibinfo {pages} {1} (\bibinfo {year} {2022})}\BibitemShut
  {NoStop}%
\bibitem [{\citenamefont {Beardo}\ \emph {et~al.}(2021)\citenamefont {Beardo},
  \citenamefont {L{\'o}pez-Su{\'a}rez}, \citenamefont {P{\'e}rez},
  \citenamefont {Sendra}, \citenamefont {Alonso}, \citenamefont {Melis},
  \citenamefont {Bafaluy}, \citenamefont {Camacho}, \citenamefont {Colombo},
  \citenamefont {Rurali} \emph {et~al.}}]{beardo2021observation}%
  \BibitemOpen
  \bibfield  {author} {\bibinfo {author} {\bibfnamefont {A.}~\bibnamefont
  {Beardo}}, \bibinfo {author} {\bibfnamefont {M.}~\bibnamefont
  {L{\'o}pez-Su{\'a}rez}}, \bibinfo {author} {\bibfnamefont {L.~A.}\
  \bibnamefont {P{\'e}rez}}, \bibinfo {author} {\bibfnamefont {L.}~\bibnamefont
  {Sendra}}, \bibinfo {author} {\bibfnamefont {M.~I.}\ \bibnamefont {Alonso}},
  \bibinfo {author} {\bibfnamefont {C.}~\bibnamefont {Melis}}, \bibinfo
  {author} {\bibfnamefont {J.}~\bibnamefont {Bafaluy}}, \bibinfo {author}
  {\bibfnamefont {J.}~\bibnamefont {Camacho}}, \bibinfo {author} {\bibfnamefont
  {L.}~\bibnamefont {Colombo}}, \bibinfo {author} {\bibfnamefont
  {R.}~\bibnamefont {Rurali}}, \emph {et~al.},\ }\bibfield  {title} {\bibinfo
  {title} {{Observation of second sound in a rapidly varying temperature field
  in Ge}},\ }\href@noop {} {\bibfield  {journal} {\bibinfo  {journal} {Sci.
  Adv.}\ }\textbf {\bibinfo {volume} {7}},\ \bibinfo {pages} {eabg4677}
  (\bibinfo {year} {2021})}\BibitemShut {NoStop}%
\bibitem [{\citenamefont {Jackson}\ \emph {et~al.}(1970)\citenamefont
  {Jackson}, \citenamefont {Walker},\ and\ \citenamefont
  {McNelly}}]{jackson1970second}%
  \BibitemOpen
  \bibfield  {author} {\bibinfo {author} {\bibfnamefont {H.~E.}\ \bibnamefont
  {Jackson}}, \bibinfo {author} {\bibfnamefont {C.~T.}\ \bibnamefont
  {Walker}},\ and\ \bibinfo {author} {\bibfnamefont {T.~F.}\ \bibnamefont
  {McNelly}},\ }\bibfield  {title} {\bibinfo {title} {{Second sound in NaF}},\
  }\href@noop {} {\bibfield  {journal} {\bibinfo  {journal} {Phys. Rev. Lett.}\
  }\textbf {\bibinfo {volume} {25}},\ \bibinfo {pages} {26} (\bibinfo {year}
  {1970})}\BibitemShut {NoStop}%
\bibitem [{\citenamefont {Narayanamurti}\ and\ \citenamefont
  {Dynes}(1972)}]{narayanamurti1972observation}%
  \BibitemOpen
  \bibfield  {author} {\bibinfo {author} {\bibfnamefont {V.}~\bibnamefont
  {Narayanamurti}}\ and\ \bibinfo {author} {\bibfnamefont {R.}~\bibnamefont
  {Dynes}},\ }\bibfield  {title} {\bibinfo {title} {Observation of second sound
  in bismuth},\ }\href@noop {} {\bibfield  {journal} {\bibinfo  {journal}
  {Phys. Rev. Lett.}\ }\textbf {\bibinfo {volume} {28}},\ \bibinfo {pages}
  {1461} (\bibinfo {year} {1972})}\BibitemShut {NoStop}%
\bibitem [{\citenamefont {Osborne}(1951)}]{osborne1951second}%
  \BibitemOpen
  \bibfield  {author} {\bibinfo {author} {\bibfnamefont {D.}~\bibnamefont
  {Osborne}},\ }\bibfield  {title} {\bibinfo {title} {{Second sound in liquid
  helium II}},\ }\href@noop {} {\bibfield  {journal} {\bibinfo  {journal}
  {Proc. Phys. Soc. A}\ }\textbf {\bibinfo {volume} {64}},\ \bibinfo {pages}
  {114} (\bibinfo {year} {1951})}\BibitemShut {NoStop}%
\bibitem [{\citenamefont {Callaway}(1959)}]{callaway1959model}%
  \BibitemOpen
  \bibfield  {author} {\bibinfo {author} {\bibfnamefont {J.}~\bibnamefont
  {Callaway}},\ }\bibfield  {title} {\bibinfo {title} {Model for lattice
  thermal conductivity at low temperatures},\ }\href@noop {} {\bibfield
  {journal} {\bibinfo  {journal} {Phys. Rev.}\ }\textbf {\bibinfo {volume}
  {113}},\ \bibinfo {pages} {1046} (\bibinfo {year} {1959})}\BibitemShut
  {NoStop}%
\bibitem [{\citenamefont {Guyer}\ and\ \citenamefont
  {Krumhansl}(1966)}]{guyer1966thermal}%
  \BibitemOpen
  \bibfield  {author} {\bibinfo {author} {\bibfnamefont {R.}~\bibnamefont
  {Guyer}}\ and\ \bibinfo {author} {\bibfnamefont {J.}~\bibnamefont
  {Krumhansl}},\ }\bibfield  {title} {\bibinfo {title} {Thermal conductivity,
  second sound, and phonon hydrodynamic phenomena in nonmetallic crystals},\
  }\href@noop {} {\bibfield  {journal} {\bibinfo  {journal} {Phys. Rev.}\
  }\textbf {\bibinfo {volume} {148}},\ \bibinfo {pages} {778} (\bibinfo {year}
  {1966})}\BibitemShut {NoStop}%
\bibitem [{\citenamefont {Ward}\ and\ \citenamefont
  {Wilks}(1951)}]{ward1951velocity}%
  \BibitemOpen
  \bibfield  {author} {\bibinfo {author} {\bibfnamefont {J.}~\bibnamefont
  {Ward}}\ and\ \bibinfo {author} {\bibfnamefont {J.}~\bibnamefont {Wilks}},\
  }\bibfield  {title} {\bibinfo {title} {The velocity of second sound in liquid
  helium near the absolute zero},\ }\href@noop {} {\bibfield  {journal}
  {\bibinfo  {journal} {Lond. Edinb. Dublin philos. mag. j. sci.}\ }\textbf
  {\bibinfo {volume} {42}},\ \bibinfo {pages} {314} (\bibinfo {year}
  {1951})}\BibitemShut {NoStop}%
\bibitem [{\citenamefont {Dingle}\ and\ \citenamefont
  {B}(1952)}]{Dingle1952Derivation}%
  \BibitemOpen
  \bibfield  {author} {\bibinfo {author} {\bibnamefont {Dingle}}\ and\ \bibinfo
  {author} {\bibfnamefont {R.}~\bibnamefont {B}},\ }\bibfield  {title}
  {\bibinfo {title} {Derivation of the velocity of second sound from maxwell's
  equation of transfer},\ }\href@noop {} {\bibfield  {journal} {\bibinfo
  {journal} {Proc. Phys. Soc. A}\ }\textbf {\bibinfo {volume} {65}},\ \bibinfo
  {pages} {374} (\bibinfo {year} {1952})}\BibitemShut {NoStop}%
\bibitem [{\citenamefont {Kwok}(1967)}]{kwok1967dispersion}%
  \BibitemOpen
  \bibfield  {author} {\bibinfo {author} {\bibfnamefont {P.~C.}\ \bibnamefont
  {Kwok}},\ }\bibfield  {title} {\bibinfo {title} {Dispersion and damping of
  second sound in non-isotropic solids},\ }\href@noop {} {\bibfield  {journal}
  {\bibinfo  {journal} {Physics Physique Fizika}\ }\textbf {\bibinfo {volume}
  {3}},\ \bibinfo {pages} {221} (\bibinfo {year} {1967})}\BibitemShut {NoStop}%
\bibitem [{\citenamefont {Sussmann}\ and\ \citenamefont
  {Thellung}(1963)}]{sussmann1963thermal}%
  \BibitemOpen
  \bibfield  {author} {\bibinfo {author} {\bibfnamefont {J.}~\bibnamefont
  {Sussmann}}\ and\ \bibinfo {author} {\bibfnamefont {A.}~\bibnamefont
  {Thellung}},\ }\bibfield  {title} {\bibinfo {title} {Thermal conductivity of
  perfect dielectric crystals in the absence of umklapp processes},\
  }\href@noop {} {\bibfield  {journal} {\bibinfo  {journal} {Proc. Phys. Soc.}\
  }\textbf {\bibinfo {volume} {81}},\ \bibinfo {pages} {1122} (\bibinfo {year}
  {1963})}\BibitemShut {NoStop}%
\bibitem [{\citenamefont {Beck}\ \emph {et~al.}(1974)\citenamefont {Beck},
  \citenamefont {Meier},\ and\ \citenamefont {Thellung}}]{beck1974phonon}%
  \BibitemOpen
  \bibfield  {author} {\bibinfo {author} {\bibfnamefont {H.}~\bibnamefont
  {Beck}}, \bibinfo {author} {\bibfnamefont {P.}~\bibnamefont {Meier}},\ and\
  \bibinfo {author} {\bibfnamefont {A.}~\bibnamefont {Thellung}},\ }\bibfield
  {title} {\bibinfo {title} {Phonon hydrodynamics in solids},\ }\href@noop {}
  {\bibfield  {journal} {\bibinfo  {journal} {Physica Status Solidi (a)}\
  }\textbf {\bibinfo {volume} {24}},\ \bibinfo {pages} {11} (\bibinfo {year}
  {1974})}\BibitemShut {NoStop}%
\bibitem [{\citenamefont {Chester}(1963)}]{chester1963second}%
  \BibitemOpen
  \bibfield  {author} {\bibinfo {author} {\bibfnamefont {M.}~\bibnamefont
  {Chester}},\ }\bibfield  {title} {\bibinfo {title} {Second sound in solids},\
  }\href@noop {} {\bibfield  {journal} {\bibinfo  {journal} {Phys. Rev.}\
  }\textbf {\bibinfo {volume} {131}},\ \bibinfo {pages} {2013} (\bibinfo {year}
  {1963})}\BibitemShut {NoStop}%
\bibitem [{\citenamefont {Prohofsky}\ and\ \citenamefont
  {Krumhansl}(1964)}]{prohofsky1964second}%
  \BibitemOpen
  \bibfield  {author} {\bibinfo {author} {\bibfnamefont {E.}~\bibnamefont
  {Prohofsky}}\ and\ \bibinfo {author} {\bibfnamefont {J.}~\bibnamefont
  {Krumhansl}},\ }\bibfield  {title} {\bibinfo {title} {Second-sound
  propagation in dielectric solids},\ }\href@noop {} {\bibfield  {journal}
  {\bibinfo  {journal} {Phys. Rev.}\ }\textbf {\bibinfo {volume} {133}},\
  \bibinfo {pages} {A1403} (\bibinfo {year} {1964})}\BibitemShut {NoStop}%
\bibitem [{\citenamefont {Li}\ \emph {et~al.}(2022)\citenamefont {Li},
  \citenamefont {Luo}, \citenamefont {Wang}, \citenamefont {Xie}, \citenamefont
  {Liu}, \citenamefont {Hu}, \citenamefont {Chen}, \citenamefont {Yao},\ and\
  \citenamefont {Pan}}]{li2022second}%
  \BibitemOpen
  \bibfield  {author} {\bibinfo {author} {\bibfnamefont {X.}~\bibnamefont
  {Li}}, \bibinfo {author} {\bibfnamefont {X.}~\bibnamefont {Luo}}, \bibinfo
  {author} {\bibfnamefont {S.}~\bibnamefont {Wang}}, \bibinfo {author}
  {\bibfnamefont {K.}~\bibnamefont {Xie}}, \bibinfo {author} {\bibfnamefont
  {X.-P.}\ \bibnamefont {Liu}}, \bibinfo {author} {\bibfnamefont
  {H.}~\bibnamefont {Hu}}, \bibinfo {author} {\bibfnamefont {Y.-A.}\
  \bibnamefont {Chen}}, \bibinfo {author} {\bibfnamefont {X.-C.}\ \bibnamefont
  {Yao}},\ and\ \bibinfo {author} {\bibfnamefont {J.-W.}\ \bibnamefont {Pan}},\
  }\bibfield  {title} {\bibinfo {title} {Second sound attenuation near quantum
  criticality},\ }\href@noop {} {\bibfield  {journal} {\bibinfo  {journal}
  {Science}\ }\textbf {\bibinfo {volume} {375}},\ \bibinfo {pages} {528}
  (\bibinfo {year} {2022})}\BibitemShut {NoStop}%
\bibitem [{\citenamefont {Baudoin}\ and\ \citenamefont
  {Thomas}(2020)}]{baudoin2020acoustic}%
  \BibitemOpen
  \bibfield  {author} {\bibinfo {author} {\bibfnamefont {M.}~\bibnamefont
  {Baudoin}}\ and\ \bibinfo {author} {\bibfnamefont {J.-L.}\ \bibnamefont
  {Thomas}},\ }\bibfield  {title} {\bibinfo {title} {Acoustic tweezers for
  particle and fluid micromanipulation},\ }\href@noop {} {\bibfield  {journal}
  {\bibinfo  {journal} {Ann. Rev. Fluid Mech.}\ }\textbf {\bibinfo {volume}
  {52}},\ \bibinfo {pages} {205} (\bibinfo {year} {2020})}\BibitemShut
  {NoStop}%
\bibitem [{\citenamefont {Sapozhnikov}\ and\ \citenamefont
  {Bailey}(2013)}]{sapozhnikov2013radiation}%
  \BibitemOpen
  \bibfield  {author} {\bibinfo {author} {\bibfnamefont {O.~A.}\ \bibnamefont
  {Sapozhnikov}}\ and\ \bibinfo {author} {\bibfnamefont {M.~R.}\ \bibnamefont
  {Bailey}},\ }\bibfield  {title} {\bibinfo {title} {Radiation force of an
  arbitrary acoustic beam on an elastic sphere in a fluid},\ }\href@noop {}
  {\bibfield  {journal} {\bibinfo  {journal} {J. Acoust. Soc. Am.}\ }\textbf
  {\bibinfo {volume} {133}},\ \bibinfo {pages} {661} (\bibinfo {year}
  {2013})}\BibitemShut {NoStop}%
\bibitem [{\citenamefont {Gong}\ and\ \citenamefont
  {Baudoin}(2020)}]{gong2020acoustic}%
  \BibitemOpen
  \bibfield  {author} {\bibinfo {author} {\bibfnamefont {Z.}~\bibnamefont
  {Gong}}\ and\ \bibinfo {author} {\bibfnamefont {M.}~\bibnamefont {Baudoin}},\
  }\bibfield  {title} {\bibinfo {title} {Acoustic radiation torque on a
  particle in a fluid: An angular spectrum based compact expression},\
  }\href@noop {} {\bibfield  {journal} {\bibinfo  {journal} {J. Acoust. Soc.
  Am.}\ }\textbf {\bibinfo {volume} {148}},\ \bibinfo {pages} {3131} (\bibinfo
  {year} {2020})}\BibitemShut {NoStop}%
\bibitem [{\citenamefont {Silva}(2011)}]{silva2011expression}%
  \BibitemOpen
  \bibfield  {author} {\bibinfo {author} {\bibfnamefont {G.~T.}\ \bibnamefont
  {Silva}},\ }\bibfield  {title} {\bibinfo {title} {{An expression for the
  radiation force exerted by an acoustic beam with arbitrary wavefront (L)}},\
  }\href@noop {} {\bibfield  {journal} {\bibinfo  {journal} {J. Acoust. Soc.
  Am.}\ }\textbf {\bibinfo {volume} {130}},\ \bibinfo {pages} {3541} (\bibinfo
  {year} {2011})}\BibitemShut {NoStop}%
\bibitem [{\citenamefont {Baresch}\ \emph {et~al.}(2013)\citenamefont
  {Baresch}, \citenamefont {Thomas},\ and\ \citenamefont
  {Marchiano}}]{baresch2013three}%
  \BibitemOpen
  \bibfield  {author} {\bibinfo {author} {\bibfnamefont {D.}~\bibnamefont
  {Baresch}}, \bibinfo {author} {\bibfnamefont {J.-L.}\ \bibnamefont
  {Thomas}},\ and\ \bibinfo {author} {\bibfnamefont {R.}~\bibnamefont
  {Marchiano}},\ }\bibfield  {title} {\bibinfo {title} {Three-dimensional
  acoustic radiation force on an arbitrarily located elastic sphere},\
  }\href@noop {} {\bibfield  {journal} {\bibinfo  {journal} {J. Acoust. Soc.
  Am.}\ }\textbf {\bibinfo {volume} {133}},\ \bibinfo {pages} {25} (\bibinfo
  {year} {2013})}\BibitemShut {NoStop}%
\bibitem [{\citenamefont {Silva}\ \emph {et~al.}(2012)\citenamefont {Silva},
  \citenamefont {Lobo},\ and\ \citenamefont {Mitri}}]{silva2012radiation}%
  \BibitemOpen
  \bibfield  {author} {\bibinfo {author} {\bibfnamefont {G.}~\bibnamefont
  {Silva}}, \bibinfo {author} {\bibfnamefont {T.}~\bibnamefont {Lobo}},\ and\
  \bibinfo {author} {\bibfnamefont {F.}~\bibnamefont {Mitri}},\ }\bibfield
  {title} {\bibinfo {title} {Radiation torque produced by an arbitrary acoustic
  wave},\ }\href@noop {} {\bibfield  {journal} {\bibinfo  {journal} {EPL}\
  }\textbf {\bibinfo {volume} {97}},\ \bibinfo {pages} {54003} (\bibinfo {year}
  {2012})}\BibitemShut {NoStop}%
\bibitem [{\citenamefont {Gong}\ \emph {et~al.}(2019)\citenamefont {Gong},
  \citenamefont {Marston},\ and\ \citenamefont {Li}}]{gong2019reversals}%
  \BibitemOpen
  \bibfield  {author} {\bibinfo {author} {\bibfnamefont {Z.}~\bibnamefont
  {Gong}}, \bibinfo {author} {\bibfnamefont {P.~L.}\ \bibnamefont {Marston}},\
  and\ \bibinfo {author} {\bibfnamefont {W.}~\bibnamefont {Li}},\ }\bibfield
  {title} {\bibinfo {title} {{Reversals of acoustic radiation torque in Bessel
  beams using theoretical and numerical implementations in three dimensions}},\
  }\href@noop {} {\bibfield  {journal} {\bibinfo  {journal} {Phys. Rev. Appl.}\
  }\textbf {\bibinfo {volume} {11}},\ \bibinfo {pages} {064022} (\bibinfo
  {year} {2019})}\BibitemShut {NoStop}%
\bibitem [{\citenamefont {Gong}\ and\ \citenamefont
  {Baudoin}(2021)}]{gong2021equivalence}%
  \BibitemOpen
  \bibfield  {author} {\bibinfo {author} {\bibfnamefont {Z.}~\bibnamefont
  {Gong}}\ and\ \bibinfo {author} {\bibfnamefont {M.}~\bibnamefont {Baudoin}},\
  }\bibfield  {title} {\bibinfo {title} {Equivalence between angular
  spectrum-based and multipole expansion-based formulas of the acoustic
  radiation force and torque},\ }\href@noop {} {\bibfield  {journal} {\bibinfo
  {journal} {J. Acoust. Soc. Am.}\ }\textbf {\bibinfo {volume} {149}},\
  \bibinfo {pages} {3469} (\bibinfo {year} {2021})}\BibitemShut {NoStop}%
\bibitem [{\citenamefont {Marston}(2006)}]{marston2006axial}%
  \BibitemOpen
  \bibfield  {author} {\bibinfo {author} {\bibfnamefont {P.~L.}\ \bibnamefont
  {Marston}},\ }\bibfield  {title} {\bibinfo {title} {Axial radiation force of
  a bessel beam on a sphere and direction reversal of the force},\ }\href@noop
  {} {\bibfield  {journal} {\bibinfo  {journal} {J. Acoust. Soc. Am.}\ }\textbf
  {\bibinfo {volume} {120}},\ \bibinfo {pages} {3518} (\bibinfo {year}
  {2006})}\BibitemShut {NoStop}%
\bibitem [{\citenamefont {Marston}(2008)}]{2008Negative}%
  \BibitemOpen
  \bibfield  {author} {\bibinfo {author} {\bibfnamefont {P.~L.}\ \bibnamefont
  {Marston}},\ }\bibfield  {title} {\bibinfo {title} {Negative axial radiation
  forces on solid spheres and shells in a bessel beam},\ }\href@noop {}
  {\bibfield  {journal} {\bibinfo  {journal} {J. Acoust. Soc. Am.}\ }\textbf
  {\bibinfo {volume} {122}},\ \bibinfo {pages} {3162} (\bibinfo {year}
  {2008})}\BibitemShut {NoStop}%
\bibitem [{\citenamefont {Mitri}(2009{\natexlab{a}})}]{mitri2009langevin}%
  \BibitemOpen
  \bibfield  {author} {\bibinfo {author} {\bibfnamefont {F.~G.}\ \bibnamefont
  {Mitri}},\ }\bibfield  {title} {\bibinfo {title} {Langevin acoustic radiation
  force of a high-order bessel beam on a rigid sphere},\ }\href@noop {}
  {\bibfield  {journal} {\bibinfo  {journal} {IEEE Trans. Ultrason.
  Ferroelectr. Freq. Control}\ }\textbf {\bibinfo {volume} {56}},\ \bibinfo
  {pages} {1059} (\bibinfo {year} {2009}{\natexlab{a}})}\BibitemShut {NoStop}%
\bibitem [{\citenamefont {Marston}\ and\ \citenamefont
  {Philip}(2009)}]{Marston2009Radiation}%
  \BibitemOpen
  \bibfield  {author} {\bibinfo {author} {\bibnamefont {Marston}}\ and\
  \bibinfo {author} {\bibfnamefont {L.}~\bibnamefont {Philip}},\ }\bibfield
  {title} {\bibinfo {title} {Radiation force of a helicoidal bessel beam on a
  sphere},\ }\href@noop {} {\bibfield  {journal} {\bibinfo  {journal} {J.
  Acoust. Soc. Am.}\ }\textbf {\bibinfo {volume} {125}},\ \bibinfo {pages}
  {3539} (\bibinfo {year} {2009})}\BibitemShut {NoStop}%
\bibitem [{\citenamefont {Zhang}\ and\ \citenamefont
  {Marston}(2011)}]{zhang2011geometrical}%
  \BibitemOpen
  \bibfield  {author} {\bibinfo {author} {\bibfnamefont {L.}~\bibnamefont
  {Zhang}}\ and\ \bibinfo {author} {\bibfnamefont {P.~L.}\ \bibnamefont
  {Marston}},\ }\bibfield  {title} {\bibinfo {title} {{Geometrical
  interpretation of negative radiation forces of acoustical Bessel beams on
  spheres}},\ }\href@noop {} {\bibfield  {journal} {\bibinfo  {journal} {Phys.
  Rev. E}\ }\textbf {\bibinfo {volume} {84}},\ \bibinfo {pages} {035601(R)}
  (\bibinfo {year} {2011})}\BibitemShut {NoStop}%
\bibitem [{\citenamefont {Gong}\ \emph {et~al.}(2021)\citenamefont {Gong},
  \citenamefont {Qiao}, \citenamefont {Fei}, \citenamefont {Li}, \citenamefont
  {Liu}, \citenamefont {Mao}, \citenamefont {He},\ and\ \citenamefont
  {Liu}}]{gong2021non}%
  \BibitemOpen
  \bibfield  {author} {\bibinfo {author} {\bibfnamefont {M.}~\bibnamefont
  {Gong}}, \bibinfo {author} {\bibfnamefont {Y.}~\bibnamefont {Qiao}}, \bibinfo
  {author} {\bibfnamefont {Z.}~\bibnamefont {Fei}}, \bibinfo {author}
  {\bibfnamefont {Y.}~\bibnamefont {Li}}, \bibinfo {author} {\bibfnamefont
  {J.}~\bibnamefont {Liu}}, \bibinfo {author} {\bibfnamefont {Y.}~\bibnamefont
  {Mao}}, \bibinfo {author} {\bibfnamefont {A.}~\bibnamefont {He}},\ and\
  \bibinfo {author} {\bibfnamefont {X.}~\bibnamefont {Liu}},\ }\bibfield
  {title} {\bibinfo {title} {Non-diffractive acoustic beams produce negative
  radiation force in certain regions},\ }\href@noop {} {\bibfield  {journal}
  {\bibinfo  {journal} {AIP Adv.}\ }\textbf {\bibinfo {volume} {11}},\ \bibinfo
  {pages} {065029} (\bibinfo {year} {2021})}\BibitemShut {NoStop}%
\bibitem [{\citenamefont {Fan}\ and\ \citenamefont
  {Zhang}(2021)}]{fan2021phase}%
  \BibitemOpen
  \bibfield  {author} {\bibinfo {author} {\bibfnamefont {X.-D.}\ \bibnamefont
  {Fan}}\ and\ \bibinfo {author} {\bibfnamefont {L.}~\bibnamefont {Zhang}},\
  }\bibfield  {title} {\bibinfo {title} {Phase shift approach for engineering
  desired radiation force: Acoustic pulling force example},\ }\href@noop {}
  {\bibfield  {journal} {\bibinfo  {journal} {J. Acoust. Soc. Am.}\ }\textbf
  {\bibinfo {volume} {150}},\ \bibinfo {pages} {102} (\bibinfo {year}
  {2021})}\BibitemShut {NoStop}%
\bibitem [{\citenamefont {Hasegawa}(1979)}]{hasegawa1979acoustic}%
  \BibitemOpen
  \bibfield  {author} {\bibinfo {author} {\bibfnamefont {T.}~\bibnamefont
  {Hasegawa}},\ }\bibfield  {title} {\bibinfo {title} {Acoustic radiation force
  on a sphere in a quasistationary wave field—theory},\ }\href@noop {}
  {\bibfield  {journal} {\bibinfo  {journal} {J. Acoust. Soc. Am.}\ }\textbf
  {\bibinfo {volume} {65}},\ \bibinfo {pages} {32} (\bibinfo {year}
  {1979})}\BibitemShut {NoStop}%
\bibitem [{\citenamefont {Mitri}(2008)}]{mitri2008acoustic}%
  \BibitemOpen
  \bibfield  {author} {\bibinfo {author} {\bibfnamefont {F.}~\bibnamefont
  {Mitri}},\ }\bibfield  {title} {\bibinfo {title} {Acoustic radiation force on
  a sphere in standing and quasi-standing zero-order bessel beam tweezers},\
  }\href@noop {} {\bibfield  {journal} {\bibinfo  {journal} {Ann. Phys.}\
  }\textbf {\bibinfo {volume} {323}},\ \bibinfo {pages} {1604} (\bibinfo {year}
  {2008})}\BibitemShut {NoStop}%
\bibitem [{\citenamefont {Mitri}(2009{\natexlab{b}})}]{mitri2009acoustic}%
  \BibitemOpen
  \bibfield  {author} {\bibinfo {author} {\bibfnamefont {F.}~\bibnamefont
  {Mitri}},\ }\bibfield  {title} {\bibinfo {title} {Acoustic radiation force of
  high-order bessel beam standing wave tweezers on a rigid sphere},\
  }\href@noop {} {\bibfield  {journal} {\bibinfo  {journal} {Ultrasonics}\
  }\textbf {\bibinfo {volume} {49}},\ \bibinfo {pages} {794} (\bibinfo {year}
  {2009}{\natexlab{b}})}\BibitemShut {NoStop}%
\bibitem [{\citenamefont {Sorbello}(1972)}]{sorbello1972phonon}%
  \BibitemOpen
  \bibfield  {author} {\bibinfo {author} {\bibfnamefont {R.~S.}\ \bibnamefont
  {Sorbello}},\ }\bibfield  {title} {\bibinfo {title} {{Phonon-Radiation Force
  in Defect Crystal Lattices}},\ }\href@noop {} {\bibfield  {journal} {\bibinfo
   {journal} {Phys. Rev. B}\ }\textbf {\bibinfo {volume} {6}},\ \bibinfo
  {pages} {4757} (\bibinfo {year} {1972})}\BibitemShut {NoStop}%
\bibitem [{\citenamefont {Gerl}(1971)}]{gerl1971calculation}%
  \BibitemOpen
  \bibfield  {author} {\bibinfo {author} {\bibfnamefont {M.}~\bibnamefont
  {Gerl}},\ }\bibfield  {title} {\bibinfo {title} {{Calculation of the Force
  Acting on an Impurity in a Metal Submitted to an Electric Field or a
  Temperature Gradient}},\ }\href@noop {} {\bibfield  {journal} {\bibinfo
  {journal} {Z. Naturforsch. A}\ }\textbf {\bibinfo {volume} {26}},\ \bibinfo
  {pages} {1} (\bibinfo {year} {1971})}\BibitemShut {NoStop}%
\bibitem [{\citenamefont {Karlsen}\ \emph {et~al.}(2016)\citenamefont
  {Karlsen}, \citenamefont {Augustsson},\ and\ \citenamefont
  {Bruus}}]{karlsen2016acoustic}%
  \BibitemOpen
  \bibfield  {author} {\bibinfo {author} {\bibfnamefont {J.~T.}\ \bibnamefont
  {Karlsen}}, \bibinfo {author} {\bibfnamefont {P.}~\bibnamefont
  {Augustsson}},\ and\ \bibinfo {author} {\bibfnamefont {H.}~\bibnamefont
  {Bruus}},\ }\bibfield  {title} {\bibinfo {title} {Acoustic force density
  acting on inhomogeneous fluids in acoustic fields},\ }\href@noop {}
  {\bibfield  {journal} {\bibinfo  {journal} {Phys. Rev. Lett.}\ }\textbf
  {\bibinfo {volume} {117}},\ \bibinfo {pages} {114504} (\bibinfo {year}
  {2016})}\BibitemShut {NoStop}%
\bibitem [{\citenamefont {Gao}\ \emph {et~al.}(2007)\citenamefont {Gao},
  \citenamefont {Jian}, \citenamefont {Zhang},\ and\ \citenamefont
  {Huang}}]{gao2007magnetophoresis}%
  \BibitemOpen
  \bibfield  {author} {\bibinfo {author} {\bibfnamefont {Y.}~\bibnamefont
  {Gao}}, \bibinfo {author} {\bibfnamefont {Y.}~\bibnamefont {Jian}}, \bibinfo
  {author} {\bibfnamefont {L.}~\bibnamefont {Zhang}},\ and\ \bibinfo {author}
  {\bibfnamefont {J.}~\bibnamefont {Huang}},\ }\bibfield  {title} {\bibinfo
  {title} {Magnetophoresis of nonmagnetic particles in ferrofluids},\
  }\href@noop {} {\bibfield  {journal} {\bibinfo  {journal} {J. Phys. Chem. C}\
  }\textbf {\bibinfo {volume} {111}},\ \bibinfo {pages} {10785} (\bibinfo
  {year} {2007})}\BibitemShut {NoStop}%
\bibitem [{\citenamefont {Xu}\ \emph {et~al.}(2018)\citenamefont {Xu},
  \citenamefont {Yang},\ and\ \citenamefont {Huang}}]{xu2018thermal1}%
  \BibitemOpen
  \bibfield  {author} {\bibinfo {author} {\bibfnamefont {L.}~\bibnamefont
  {Xu}}, \bibinfo {author} {\bibfnamefont {S.}~\bibnamefont {Yang}},\ and\
  \bibinfo {author} {\bibfnamefont {J.}~\bibnamefont {Huang}},\ }\bibfield
  {title} {\bibinfo {title} {Thermal theory for heterogeneously architected
  structure: fundamentals and application},\ }\href@noop {} {\bibfield
  {journal} {\bibinfo  {journal} {Phys. Rev. E}\ }\textbf {\bibinfo {volume}
  {98}},\ \bibinfo {pages} {052128} (\bibinfo {year} {2018})}\BibitemShut
  {NoStop}%
\bibitem [{\citenamefont {Dong}\ \emph {et~al.}(2004)\citenamefont {Dong},
  \citenamefont {Huang}, \citenamefont {Yu},\ and\ \citenamefont
  {Gu}}]{dong2004dielectric}%
  \BibitemOpen
  \bibfield  {author} {\bibinfo {author} {\bibfnamefont {L.}~\bibnamefont
  {Dong}}, \bibinfo {author} {\bibfnamefont {J.}~\bibnamefont {Huang}},
  \bibinfo {author} {\bibfnamefont {K.}~\bibnamefont {Yu}},\ and\ \bibinfo
  {author} {\bibfnamefont {G.}~\bibnamefont {Gu}},\ }\bibfield  {title}
  {\bibinfo {title} {Dielectric response of graded spherical particles of
  anisotropic materials},\ }\href@noop {} {\bibfield  {journal} {\bibinfo
  {journal} {J. Appl. Phys.}\ }\textbf {\bibinfo {volume} {95}},\ \bibinfo
  {pages} {621} (\bibinfo {year} {2004})}\BibitemShut {NoStop}%
\bibitem [{\citenamefont {Huang}\ \emph {et~al.}(2003)\citenamefont {Huang},
  \citenamefont {Karttunen}, \citenamefont {Yu},\ and\ \citenamefont
  {Dong}}]{huang2003dielectrophoresis}%
  \BibitemOpen
  \bibfield  {author} {\bibinfo {author} {\bibfnamefont {J.}~\bibnamefont
  {Huang}}, \bibinfo {author} {\bibfnamefont {M.}~\bibnamefont {Karttunen}},
  \bibinfo {author} {\bibfnamefont {K.}~\bibnamefont {Yu}},\ and\ \bibinfo
  {author} {\bibfnamefont {L.}~\bibnamefont {Dong}},\ }\bibfield  {title}
  {\bibinfo {title} {Dielectrophoresis of charged colloidal suspensions},\
  }\href@noop {} {\bibfield  {journal} {\bibinfo  {journal} {Phys. Rev. E}\
  }\textbf {\bibinfo {volume} {67}},\ \bibinfo {pages} {021403} (\bibinfo
  {year} {2003})}\BibitemShut {NoStop}%
\bibitem [{\citenamefont {Ye}\ and\ \citenamefont {Huang}(2008)}]{ye2008non}%
  \BibitemOpen
  \bibfield  {author} {\bibinfo {author} {\bibfnamefont {C.}~\bibnamefont
  {Ye}}\ and\ \bibinfo {author} {\bibfnamefont {J.}~\bibnamefont {Huang}},\
  }\bibfield  {title} {\bibinfo {title} {Non-classical oscillator model for
  persistent fluctuations in stock markets},\ }\href@noop {} {\bibfield
  {journal} {\bibinfo  {journal} {Physica A: Statistical Mechanics and its
  Applications}\ }\textbf {\bibinfo {volume} {387}},\ \bibinfo {pages} {1255}
  (\bibinfo {year} {2008})}\BibitemShut {NoStop}%
\bibitem [{\citenamefont {Shen}\ \emph {et~al.}(2016)\citenamefont {Shen},
  \citenamefont {Jiang}, \citenamefont {Li},\ and\ \citenamefont
  {Huang}}]{shen2016thermal1}%
  \BibitemOpen
  \bibfield  {author} {\bibinfo {author} {\bibfnamefont {X.}~\bibnamefont
  {Shen}}, \bibinfo {author} {\bibfnamefont {C.}~\bibnamefont {Jiang}},
  \bibinfo {author} {\bibfnamefont {Y.}~\bibnamefont {Li}},\ and\ \bibinfo
  {author} {\bibfnamefont {J.}~\bibnamefont {Huang}},\ }\bibfield  {title}
  {\bibinfo {title} {Thermal metamaterial for convergent transfer of conductive
  heat with high efficiency},\ }\href@noop {} {\bibfield  {journal} {\bibinfo
  {journal} {Appl. Phys. Lett.}\ }\textbf {\bibinfo {volume} {109}} (\bibinfo
  {year} {2016})}\BibitemShut {NoStop}%
\bibitem [{\citenamefont {Liu}\ \emph {et~al.}(2013)\citenamefont {Liu},
  \citenamefont {Wei}, \citenamefont {Zhang}, \citenamefont {Xin},\ and\
  \citenamefont {Huang}}]{liu2013statistical}%
  \BibitemOpen
  \bibfield  {author} {\bibinfo {author} {\bibfnamefont {L.}~\bibnamefont
  {Liu}}, \bibinfo {author} {\bibfnamefont {J.}~\bibnamefont {Wei}}, \bibinfo
  {author} {\bibfnamefont {H.}~\bibnamefont {Zhang}}, \bibinfo {author}
  {\bibfnamefont {J.}~\bibnamefont {Xin}},\ and\ \bibinfo {author}
  {\bibfnamefont {J.}~\bibnamefont {Huang}},\ }\bibfield  {title} {\bibinfo
  {title} {A statistical physics view of pitch fluctuations in the classical
  music from bach to chopin: Evidence for scaling},\ }\href@noop {} {\bibfield
  {journal} {\bibinfo  {journal} {PLoS One}\ }\textbf {\bibinfo {volume} {8}},\
  \bibinfo {pages} {e58710} (\bibinfo {year} {2013})}\BibitemShut {NoStop}%
\bibitem [{\citenamefont {Huang}\ and\ \citenamefont
  {Yu}(2005)}]{huang2005magneto}%
  \BibitemOpen
  \bibfield  {author} {\bibinfo {author} {\bibfnamefont {J.}~\bibnamefont
  {Huang}}\ and\ \bibinfo {author} {\bibfnamefont {K.}~\bibnamefont {Yu}},\
  }\bibfield  {title} {\bibinfo {title} {Magneto-controlled nonlinear optical
  materials},\ }\href@noop {} {\bibfield  {journal} {\bibinfo  {journal} {Appl.
  Phys. Lett.}\ }\textbf {\bibinfo {volume} {86}} (\bibinfo {year}
  {2005})}\BibitemShut {NoStop}%
\bibitem [{\citenamefont {Qiu}\ \emph {et~al.}(2015)\citenamefont {Qiu},
  \citenamefont {Meng},\ and\ \citenamefont {Huang}}]{qiu2015nonstraight}%
  \BibitemOpen
  \bibfield  {author} {\bibinfo {author} {\bibfnamefont {T.}~\bibnamefont
  {Qiu}}, \bibinfo {author} {\bibfnamefont {X.}~\bibnamefont {Meng}},\ and\
  \bibinfo {author} {\bibfnamefont {J.}~\bibnamefont {Huang}},\ }\bibfield
  {title} {\bibinfo {title} {Nonstraight nanochannels transfer water faster
  than straight nanochannels},\ }\href@noop {} {\bibfield  {journal} {\bibinfo
  {journal} {J. Phys. Chem. B}\ }\textbf {\bibinfo {volume} {119}},\ \bibinfo
  {pages} {1496} (\bibinfo {year} {2015})}\BibitemShut {NoStop}%
\end{thebibliography}
\providecommand{\noopsort}[1]{}\providecommand{\singleletter}[1]{#1}%

\end{document}